\documentclass[preprint]{elsarticle}
\usepackage[margin=1in]{geometry}
\usepackage[utf8]{inputenc}
\usepackage{graphicx}
\usepackage{float}
\usepackage{amsmath}
\usepackage{enumitem}
\usepackage{esvect}
\usepackage[table,xcdraw]{xcolor}
\usepackage{lscape}
\usepackage{hyperref}
\usepackage[right]{lineno}
\usepackage{tikz}
\usepackage[font=small]{caption}
\usepackage{soul}
\usepackage{hyperref}
\usepackage{array}
\usepackage{amssymb}

\modulolinenumbers[5]
\journal{Journal of Theoretical Biology}

\newcommand{\blue}[1]{\textcolor{black}{#1}}

\usepackage{anyfontsize}

\begin{document}

\title{\textcolor{black}{Aging-induced fragility} of the immune system}
\author[add1]{Eric Jones\corref{fn1}}
\author[add2]{Jiming Sheng\corref{fn1}}
\author[add1]{Jean Carlson\corref{fn2}}
\author[add2]{Shenshen Wang\corref{fn2}}
\address[add1]{Department of Physics, University of California, Santa Barbara, California 93106, USA}
\address[add2]{Department of Physics \& Astronomy, University of California, Los Angeles, California 90095, USA}
\cortext[fn1]{equal contribution}
\cortext[fn2]{equal contribution}
\cortext[fn3]{ewj@physics.ucsb.edu}
\cortext[fn3]{shenshen@physics.ucla.edu}

\begin{frontmatter}
\begin{abstract}
The adaptive and innate branches of the vertebrate immune system work in close collaboration to protect organisms from harmful pathogens.
As an organism ages its immune system undergoes immunosenescence, characterized by declined performance or malfunction in either immune branch, which can lead to disease and death.
In this study we develop a \textcolor{black}{mathematical framework} of coupled innate and adaptive immune responses, namely the integrated immune branch (IIB) model.
\textcolor{black}{This model describes dynamics of immune components in both branches, uses a shape-space representation to encode pathogen-specific immune memory, and exhibits three steady states--- health, septic death, and chronic inflammation--- qualitatively similar to clinically-observed immune outcomes}.
In this model, the immune system (initialized in the health state) is subjected to a sequence of pathogen encounters, \textcolor{black}{and we use the number of prior pathogen encounters as a proxy for the ``age'' of the immune system.}
We find that repeated pathogen encounters may trigger a fragility in which any encounter with a novel pathogen will cause the system to \textcolor{black}{irreversibly switch from health to chronic inflammation}.
This transition is consistent with the onset of ``inflammaging'', a condition observed in aged individuals who experience chronic low-grade inflammation even in the absence of pathogens.
The IIB model predicts that the onset of chronic inflammation strongly depends on the history of encountered pathogens; the timing of onset differs drastically when the same set of infections occurs in a different order.
Lastly, the coupling between the innate and adaptive immune branches generates a trade-off between rapid pathogen clearance and a delayed onset of immunosenescence.
Overall, by considering the complex feedback between immune compartments, our work suggests potential mechanisms for immunosenescence and provides a theoretical framework at the system level and on the scale of an organism's lifetime to account for clinical observations.
\end{abstract}


\end{frontmatter}

\section{Introduction}

Infectious diseases as diverse as bacterial pneumonia, influenza, tuberculosis, herpes zoster, and most recently COVID-19 have an increased morbidity and mortality among the elderly \cite{yoshikawa1981important, eickhoff1961observations, powell1980rising, miller1970zoster, yoshikawa1997perspective,LiuHan2020}.
Especially in conjunction with global demographics that broadly reflect increases in age across the world’s populations (both due to prolonged life expectancy and declining birth rates), the prevalence of disease among the elderly underscores the need for a better understanding of how physiology changes with age \cite{UNDES2019tenfindings}.
In particular, it is acutely important to identify the causes of immunosenescence, the readily observed yet mechanistically vague deterioration of immune function in aged individuals.

The vertebrate immune system targets and clears pathogens through the collaborative efforts of innate and adaptive immune responses: the innate immune system reacts quickly and non-specifically to pathogenic threats, while the adaptive immune system acts more slowly and generates a pathogen-specific response through clonal expansion of cognate T and B lymphocytes.
To orchestrate this division of responsibility, extensive bidirectional interactions exist between the innate and adaptive immune compartments \cite{getz2005bridging, shanker2017innate, e2004activation, pelletier2010evidence, strutt2011control, shanker2010adaptive}.
For example, dendritic cells in the innate compartment mediate the presentation of antigens to the adaptive compartment \cite{TheryAmigorena2001}.
Conversely, T cells in the adaptive arm reduce the production of inflammatory cytokines and thus limit tissue damage caused by the innate immune response \cite{guarda2009t, palm2007not, kim2007adaptive}.
For example, experiments with nude mice (a mutant mouse strain with low T cell levels) showed that death can ensue without this adaptive suppression of inflammation \cite{palm2007not, kim2007adaptive}.

Immunosenescence manifests itself in both the innate and the adaptive immune branches.
In the adaptive branch aging is partially driven by thymus involution, which reduces the output of new naive T cells \cite{aw2011origin, goronzy2005t, shanley2009evolutionary}.
Furthermore, a lifetime of persistent pathogen exposures (e.g. chronic infections like cytomegalovirus) leads to oligoclonal expansion of memory T cells specific to those pathogens. 
These physiological mechanisms lead to an ``imbalanced'' repertoire of immune cells that is predominately populated by memory cells specific to frequently encountered pathogens, which limit the ability of the adaptive branch to respond to novel pathogens.
Indeed, the increased fraction of CD8+ T cells caused by memory inflation has been associated with the immune risk phenotype (IRP), which has been found to predict impaired immune function and mortality \cite{souquette2017constant, blackman2011narrowing, wikby2006immune}.

In the innate compartment, aging is associated with the development of a chronic low-grade inflammatory response even in the absence of pathogen stimulation, called ``inflammaging'' \cite{franceschi2000inflamm}.
The elderly often experience chronic inflammation and possess elevated levels of pro-inflammatory cytokines \cite{wikby2006immune,bruunsgaard1999elderly,bruunsgaard2003predicting,volpato2001cardiovascular}, which have been found to be strong predictors of mortality (for example, interleukin 6 has been associated with the IRP) \cite{wikby2006immune}.
Prior theories suggest that inflammaging is facilitated by long-lasting pathogen encounters, cell debris and stress, and the reduced efficiency of the adaptive immune response \cite{franceschi2000inflamm, fulop2018immunosenescence,  franceschi2018inflammaging}.
Still, the mechanisms underlying the onset of inflammaging--- and in particular its connection to aging in the adaptive immune system--- require further study.

In an earlier mathematical model of the adaptive immune response, Stromberg and Carlson found that repeated pathogen exposures could lead to an imbalanced immune repertoire that was \textcolor{black}{fragile to rare pathogens, in the sense that rare pathogens proliferated significantly more than common pathogens} \cite{StrombergCarlson2006}.
Around the same time, Reynolds {\em et al.} developed a model of the innate immune response immediately following a pathogen encounter.
Based on these earlier models, in this paper we construct an \textcolor{black}{ordinary differential equation} model of the coupled innate-adaptive immune system called the integrated immune branch (IIB) model, and demonstrate how immunosenescence can develop and trigger a chronic inflammatory response.
\textcolor{black}{As detailed in Table \ref{tab:BioComponents}, the state variables of the IIB model incorporate pathogen abundance $P_i$ (where subscript $i$ indicates the pathogen type); the inflammatory response of the innate branch with neutrophils $N^*$, anti-inflammatory cytokines $C_A$, and inflammatory tissue damage $D$; and the pathogen-specific T cell dynamics of the adaptive branch with naive cells $N_i$, memory cells $M_i$, and effector cells $E_i$.} 
Here, \textcolor{black}{the onset of immunosenescence arises purely from the fragility of an imbalanced immune repertoire shaped by past encounters of pathogens; this fragility renders the system vulnerable to novel infections, causing an irreversible transition to the chronic inflammation state.} Importantly, this fragility emerges without any assumptions regarding the degradation of cellular function with age.
The IIB model also recapitulates several clinically-observed signatures of immunosenescence: the ratio of naive to memory cells decreases over time \cite{BorenGershwin2004}, repeated exposure to chronic infections (e.g. human cytomegalovirus) induces immune fragility \cite{AielloAccardi2019}, and this fragility is characterized by chronic inflammation (``inflammaging'') \cite{BorenGershwin2004, wikby2006immune}.


With the IIB model, we first characterize the dynamics and steady states of the immune system in response to a single infection event.
\textcolor{black}{The three steady states of the IIB model--- health, septic death, and chronic inflammation--- are characterized, and their dependence on key parameters is explored. }
Then, the system is exposed to a series of pathogens that form an infection history; \textcolor{black}{these regularly-spaced pathogen encounters are used to measure the age of the immune system.}
This sequence of infection events causes overspecialization in the adaptive compartment and triggers chronic inflammation.
In particular, the order in which infections are encountered strongly influences immune outcomes, and can hasten or delay the onset of chronic inflammation.
Further, by tuning resource allocation in the adaptive compartment toward pathogen clearance versus suppression of inflammation, crosstalks between immune compartments may be directly manipulated.
This manipulation reveals a trade-off between a delayed onset of chronic inflammation and rapid pathogen clearance.
Our model provides a mechanistic explanation for how repeated pathogen exposures can cause immune fragility that leads to inflammaging and immunosenescence, and may serve as a foundation for quantitative studies of immune crosstalk and aging.


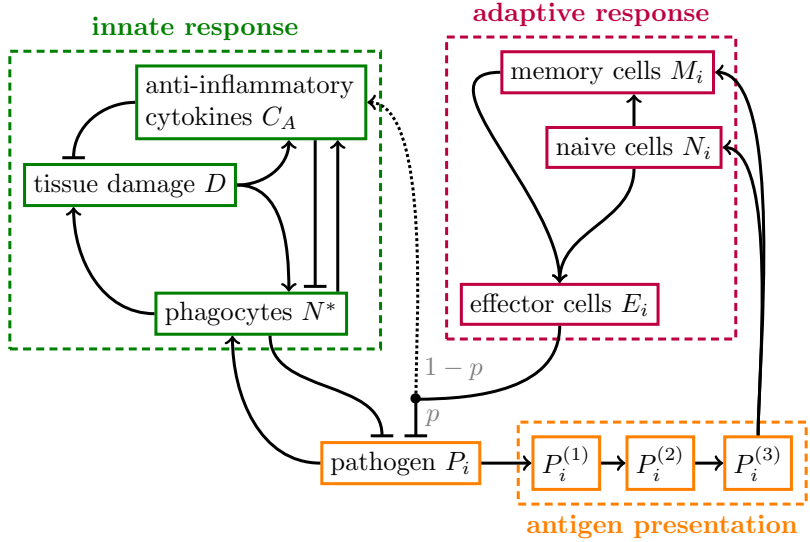
\begin{figure}[t!]
\centering
\usetikzlibrary{fit}
\begin{tikzpicture}[
    innate/.style={rectangle,draw=black!50!green,line width=0.4mm},
    adaptive/.style={rectangle,draw=purple,line width=0.4mm},
    pathogen/.style={rectangle,draw=orange,line width=0.4mm},
    helps/.style={->,line width=0.4mm},
    plain/.style={-,line width=0.4mm},
    hurts/.style={-|,line width=0.4mm}, scale = .5]

\node (xx) at (-2.6, 1.3) {\textbullet};
\node [label={[text=gray]$1-p$}] (yy) at (-1.6, 1.3) {};
\node[pathogen] (p) at (-3, -.4) {pathogen $P_i$};
\node[pathogen] (p1) at (1.4, -.4) {$P_i^{(1)}$};
\node[pathogen] (p2) at (3.9, -.4) {$P_i^{(2)}$};
\node[pathogen] (p3) at (6.5, -.4) {$P_i^{(3)}$};
\node[draw,pathogen,densely dashed,fit=(p1) (p2) (p3), inner sep=.5em,
    label={[orange, font=\bfseries]below:antigen presentation}] {};

\node[adaptive] (n) at (3.2, 8) {naive cells $N_i$};
\node[adaptive] (m) at (2.5, 10) {memory cells $M_i$};
\begin{scope}[node distance=8.5em]
\node[adaptive] (e) [above right of= p] {effector cells $E_i$};
\end{scope}
\node[draw,adaptive,densely dashed,fit=(n) (m) (e), inner sep=.5em,
    label={[purple, font=\bfseries]above:adaptive response}] {};

\begin{scope}[node distance=8em]
\node[innate] (ns) [above left of= p] {phagocytes $N^*$};
\end{scope}
\node[innate] (d) at (-10.2, 7) {tissue damage $D$};
\begin{scope}[node distance=8em]
\node[innate] (ca) [above of= ns, text width=8em] {anti-inflammatory cytokines $C_A$};
\end{scope}
\node[draw,innate,densely dashed,fit=(ns) (d) (ca), inner sep=.5em,
    label={[black!50!green, font=\bfseries]above:innate response}] {};

\draw[helps] (p) to (p1);
\draw[helps] (p1) to (p2);
\draw[helps] (p2) to (p3);
\draw[helps, looseness=.5] (p3) to[out=90, in=0] (n);
\draw[helps, looseness=.5] (p3) to[out=90, in=0] (m);
\draw[helps] (n) to[out=90, in=270] ([xshift=7mm]m.south);
\draw[helps] (n) to[out=270, in=90] (e);
\draw[helps] (m) to[out=180, in=90] (e);
\draw[plain] (e) to[out=270, in=0] (xx.center); 
\draw[hurts] (xx.center) to[out=270, in=90] node[auto,text=gray]{$p$}  ([xshift=.4cm,yshift=1mm]p.north);
\draw[helps, densely dotted, looseness=.55] (xx.center) to[out=90, in=0] (ca);
%
\draw[helps] (p) to[out=180, in=270] ([xshift=-.5cm]ns.south);
\draw[hurts] ([xshift=.5cm]ns.south) to[out=270, in=90] ([xshift=-.5cm,yshift=1mm]p.north);
\draw[helps] ([xshift=23mm]ns.north) to ([xshift=23mm]ca.south);
\draw[hurts] ([xshift=17mm]ca.south) to ([xshift=17mm,yshift=1mm]ns.north);
\draw[helps] (d)  [out=0, in=90] to ([xshift=10mm]ns.north);
\draw[helps] (d)  [out=0, in=270] to ([xshift=10mm]ca.south);
\draw[helps] (ns) [out=180, in=270] to ([xshift=-1.5cm]d.south);
\draw[hurts] (ca) [out=180,in=90] to ([xshift=-1.5cm,yshift=1mm]d.north);
\end{tikzpicture}

    \caption{{\bf Schematic of the integrated immune branch (IIB) model.}
    An introduced pathogen prompts innate and adaptive immune responses that seek to eliminate the pathogen. 
    The innate response (green) is adapted from the model of Reynolds \textit{et al}.~\cite{ReynoldsErmentrout2006}, in which the presence of a pathogen activates phagocytes that induce inflammation and the subsequent production of anti-inflammatory cells. 
    The adaptive immune response (magenta) is adapted from the model of Stromberg and Carlson \cite{StrombergCarlson2006}, in which presented pathogens (orange) activate naive and memory T cells specific to that pathogen, causing them to divide into effector cells that target the pathogen.
    \textcolor{black}{In this model, the delay in the adaptive response due to antigen presentation is hard-coded via the compartments $P_i^{(1)}$, $P_i^{(2)}$, and $P_i^{(3)}$ using the linear chain technique.}
    The state variables of this model are described in Table~\ref{tab:BioComponents}, the model itself is given explicitly in Table~\ref{tab:model_eqns}, and the model parameters are provided in Table \ref{tab:Params}. 
    \label{fig:schematic}}
\end{figure}


\begin{table}[t!]
\begin{tabular}{|c|p{3.1cm}|p{10.7cm}|}
\hline
\rowcolor[HTML]{9B9B9B} 
Notation & Immune component & Description \\ \hline
$P_i$ & Pathogen & 
\vspace*{-.8em} \begin{itemize}[noitemsep,topsep=0pt,leftmargin=*]
\item Harmful exogenous stimulants (e.g., bacteria or viruses) that \text{activate} an immune response
\item Pathogen of shape $i$ (shape space formulation) 
\end{itemize}\baselineskip 0pt \\ \hline
\rowcolor[HTML]{D9D9D9} 
$N^*$ & Activated \text{phagocytes} &
\vspace*{-.8em} \begin{itemize}[noitemsep,topsep=0pt,leftmargin=*]
\item Phagocytes (which include neutrophils and macrophages) that are anctivated by any pathogen $P_i$ or by pro-inflammatory cytokines
\item Responsible for removing pathogens $P_i$, but cause collateral tissue damage $D$ 
\end{itemize}\baselineskip 0pt \\ \hline
$D$ & Tissue damage &
\vspace*{-.8em} \begin{itemize}[noitemsep,topsep=0pt,leftmargin=*]
\item Caused by activated phagocytes $N^*$
\item Causes release of pro-inflammatory cytokines that recruit additional phagocytes $N^*$ 
\end{itemize}\baselineskip 0pt \\ \hline
\rowcolor[HTML]{D9D9D9} 
$C_A$ & Anti-inflammatory cytokines &
\vspace*{-.8em} \begin{itemize}[noitemsep,topsep=0pt,leftmargin=*]
\item Small protein molecules that reduce the efficiency and recruitment of activated phagocytes $N^*$
\item Production encouraged by activated phagocytes $N^*$ and tissue damage $D$, and also by effector cells $E$ (innate-adaptive crosstalk) 
\end{itemize}\baselineskip 0pt \\ \hline
$N_i$ & Naive cells & 
\vspace*{-.8em} \begin{itemize}[noitemsep,topsep=0pt,leftmargin=*]
\item Mature T cells with receptor specificity represented by shape $i$ that are agnostic to previous pathogen encounters
\item Divide and differentiate into memory $M_i$ and effector $E_i$ cells when activated by pathogens $P_i$
\item Subject to homeostasis control mechanisms 
\end{itemize}\baselineskip 0pt \\ \hline
\rowcolor[HTML]{D9D9D9} 
$M_i$ & Memory cells & 
\vspace*{-.8em} \begin{itemize}[noitemsep,topsep=0pt,leftmargin=*]
\item Long-lived cells differentiated from naive cells $N_i$ with the same pathogen specificity
\item Divide and differentiate into memory $M_i$ and effector $E_i$ cells when activated by pathogens $P_i$
\item Subject to homeostasis control mechanisms
\end{itemize}\baselineskip 0pt \\ \hline
$E_i$ & Effector cells & 
\vspace*{-.8em} \begin{itemize}[noitemsep,topsep=0pt,leftmargin=*]
\item Short-lived cells differentiated from naive $N_i$ and memory $M_i$ cells 
\item Remove pathogen $P_i$ and produces anti-inflammatory cytokines $C_A$ 
\end{itemize}\baselineskip 0pt \\ \hline
\end{tabular}
\caption{{\bf Major biological components in the integrated immune branch (IIB) model.} Model equations are provided in full in Table \ref{tab:model_eqns}.
\label{tab:BioComponents}}
\end{table}


\section{Mathematical model}
Extensive mathematical and computational modeling efforts have been made to better understand both the innate and adaptive branches of vertebrate immune system \cite{vodovotz2017solving, eftimie2016mathematical, chakraborty2017perspective}.
Additionally, a rich literature exists regarding the inflammatory innate response \cite{dunster2014resolution, nagaraja2014computational, vodovotz2006silico, vodovotz2009mechanistic, yang2011quantitative} and the adaptive immune repertoire \cite{de1994t, stirk2008stochastic,davis2013mathematical, schlesinger2014coevolutionary}.
In particular, Reynolds \textit{et al.} studied the positive feedback between activated phagocytes and collateral tissue damage \cite{ReynoldsErmentrout2006}, and Stromberg and Carlson modeled the accumulated loss of memory cell diversity over the course of a lifetime of infections \cite{StrombergCarlson2006}.
In this study, we modify and synthesize the models of Reynolds \textit{et al.} and Stromberg and Carlson to develop an integrated immune branch (IIB) model, in which the innate and adaptive immune branches work collaboratively to clear pathogens.

A schematic of the IIB model is depicted in Fig.~\ref{fig:schematic}, and a thorough accounting of the individual components of the innate (pathogen $P_i$, phagocytes $N^*$, tissue damage $D$, and anti-inflammatory cytokines $C_A$) and adaptive (pathogen $P_i$, naive cells $N_i$, memory cells $M_i$, and effector cells $E_i$) models is provided in Table \ref{tab:BioComponents}.
The coupled model is described in full in Table~\ref{tab:model_eqns}, and parameter descriptions and values are provided in Table \ref{tab:Params}.
For more complete descriptions of the separate innate and adaptive immune models, we refer the reader to the original publications \cite{ReynoldsErmentrout2006, StrombergCarlson2006}.

\subsection{Innate immune response}
In the model formulated by Reynolds \textit{et al}.~\cite{ReynoldsErmentrout2006} (depicted in the green box of Fig.~\ref{fig:schematic}), once a pathogen $P_i$ is introduced, phagocytes $N^*$ (which include, e.g., neutrophils and macrophages) are recruited to the site of the infection by activation or migration.
These phagocytes kill the pathogen $P_i$ by phagocytosis, degranulation, or by creating neutrophil extracellular traps.
At the same time, these phagocytes release inflammatory cytokines that induce inflammation and tissue damage $D$ in the host.
Damaged tissue in turn releases additional inflammatory cytokines, further promoting phagocyte activation.
Following this initial inflammatory response, a wave of anti-inflammatory cytokines $C_A$ is released to downregulate phagocyte recruitment and to reduce inflammation and tissue damage.

The steady-state behavior of this model was investigated by Reynolds \textit{et al}.~\cite{ReynoldsErmentrout2006}, who found that this formulation of the innate immune response led to three possible steady states: \textit{health}, in which the pathogen is cleared and the phagocytes and tissue damage vanish; \textit{septic death}, in which the innate response is unable to clear the pathogen; and \textit{chronic inflammation}, in which the innate response clears the pathogen at the expense of inducing a constant inflammatory response that fails to dissipate, even after pathogen clearance. 
In the model's original formulation this chronic inflammation steady state was named \textit{aseptic death}, but in this paper we allow for the possible interpretation of this steady state as ``inflammaging,'' which is not necessarily fatal, rather than death \cite{chen2010sterile, rock2009sterile}.
As we will show, the IIB model retains these three steady states while introducing interplay between the non-specific innate response and the specific adaptive response.

\begin{table}[t!]
{\small
\begin{tabular}{|c|p{9cm}|}
\hline
\rowcolor[HTML]{9B9B9B} 
Equation & Interpretation \\ \hline
\raisebox{-4em}{
$\begin{aligned} \frac{\text{d}P_i}{\text{d}t} &= k_{pg} P_i \left(1 - \frac{|\vv{P}|}{P_\infty} \right)
    - \frac{k_{pm} s_m}{\mu_m + k_{mp} |\vv{P}|}P_i \\ 
    &\quad - k_{pn} f(N^*, \ C_A) P_i - p\gamma P_i E_i -\Delta \, P_i\end{aligned}$}
 & 
 \vspace{-.5em}Pathogen $P_i$ of shape $i$ changes according to:
 \begin{itemize}[noitemsep,topsep=0pt,leftmargin=*]
     \item logistic growth, carrying capacity $P_\infty$ ($|\cdot|$ denotes the 1-norm)
     \item inhibition by a non-local immune response
     \item clearance by innate phagocytes $N^*$, effect is mediated by anti-inflammatory cytokines $C_A$ via $f(\cdot)$
     \item clearance by adaptive effector cells $E_i$
     \item sequestration by dendritic cells for antigen presentation \end{itemize} \baselineskip 0pt
 \\ \hline
\rowcolor[HTML]{D9D9D9} 
\raisebox{-3.9em}{$\begin{aligned}
\frac{\text{d}P_i^{(1)}}{\text{d}t} &=\Delta \left[P_i - P_i^{(1)}\right]\\
\frac{\text{d}P_i^{(2)}}{\text{d}t} &=\Delta \left[P_i^{(1)} - P_i^{(2)}\right]\\
\frac{\text{d}P_i^{(3)}}{\text{d}t} &=\Delta \left[P_i^{(2)} - P_i^{(3)} \right] - \beta \gamma P_i^{(3)} (N_i+M_i)\\
\end{aligned}$}
& \vspace{-.8em} 
\begin{itemize}[noitemsep,topsep=0pt,leftmargin=*]
\item antigen presentation occurs with hard-coded delay (linear chain technique) of $3/\Delta$ units of time on average required for a pathogen $P_i$ to transition to compartment $P_i^{(3)}$
\item compartments $P_i^{\text{(1, 2, 3)}}$ correspond to intermediate states during antigen presentation
\item once antigen arrives in compartment $P_i^{(3)}$ it activates naive cells $N_i$ and memory cells $M_i$
\end{itemize} \baselineskip 0pt
\\ \hline
$\begin{aligned}
& \\ 
\frac{\text{d}N_i}{\text{d}t} &= -\alpha \gamma N_i  P_i^{(3)} + \theta_N - \delta_N N_i \frac{|\vv{M}| + |\vv{N}|}{R_0}\\
\end{aligned}$
& \vspace{-2em}
Naive cells $N_i$ of shape $i$ change according to:
\begin{itemize}[noitemsep,topsep=0pt,leftmargin=*]
\item division into effector cells $E_i$ (with rate $f$) and memory cells $M_i$ (with rate $1-f$)
\item constant production at rate $\theta_N$ 
\item return to homeostatic equilibrium (timescale $1/\delta_N$)
\end{itemize} \baselineskip 0pt 
\\\hline
\rowcolor[HTML]{D9D9D9}
\raisebox{-3em}{
$\begin{aligned} 
\frac{\text{d}M_i}{\text{d}t} &= (2 - 2f) \alpha \gamma N_i P_i^{(3)}
    \\ &\quad + (1 - 2f) \alpha \gamma M_i P_i^{(3)}
    - \delta_M M_i \frac{|\vv{M}| + |\vv{N}|}{R_0}\\
\end{aligned}$}
& \vspace{-.8em}
Memory cells $M_i$ of shape $i$ change according to:
\begin{itemize}[noitemsep,topsep=0pt,leftmargin=*]
\item division into effector cells $E_i$ (with rate $f$) and memory cells $M_i$ (with rate $1-f$); \textcolor{black}{factor of $2$ results from cell division}
\item growth from naive and memory cell division
\item decay at rate $\delta_M$
\end{itemize}\baselineskip 0pt
\\\hline
$\begin{aligned} \\
\frac{\text{d}E_i}{\text{d}t} &= 2 f \alpha (M_i + N_i) \gamma P_i^{(3)} - \delta_E E_i \\
\end{aligned}$
& \vspace{-1.8em}
Effector cells $E_i$ of shape $i$ change according to:
\begin{itemize}[noitemsep,topsep=0pt,leftmargin=*]
\item production by naive and memory cells proportional to antigen presentation rate $\alpha$
\item decay at rate $\delta_E$
\end{itemize}\baselineskip 0pt
\\\hline
\rowcolor[HTML]{D9D9D9} 
$\begin{aligned} \\
\frac{\text{d}N^*}{\text{d}t} &= \frac{s_{nr} R}{\mu_{nr} + R} - \mu_n N^*\\
\end{aligned}$
& 
\vspace{-2em} Innate phagocytes $N^*$ change according to:
\begin{itemize}[noitemsep,topsep=0pt,leftmargin=*]
\item activation by the presence of other phagocytes, pathogen, or tissue damage (encapsulated by $R$)
\item decay at rate $\mu_n$
\end{itemize}\baselineskip 0pt
\\\hline
$\begin{aligned} \\
\frac{\text{d}D}{\text{d}t} &= k_{dn} f_s ( f(N^*,\ C_A)) - \mu_d D\\
\end{aligned}$
& 
\vspace{-1.8em} Tissue damage $D$ changes according to:
\begin{itemize}[noitemsep,topsep=0pt,leftmargin=*]
\item induced by activated phagocytes $N^*$, but ameliorated by the presence of anti-inflammatory cytokines $C_A$ via $f(\cdot)$
\item decay at rate $\mu_d$ 
\end{itemize}\baselineskip 0pt
\\\hline
\rowcolor[HTML]{D9D9D9} 
\raisebox{-2.5em}{
$\begin{aligned}
\frac{\text{d}C_A}{\text{d}t} &= s_c 
    + k_{cn} \frac{f(N^* + k_{cnd}D,\ C_A)}{1 + f(N^* + k_{cnd}D,\ C_A)} - \mu_c C_A \\
    &\quad + (1-p)\, k_{ce} \frac{|\vv{E}|}{|\vv{E}|+E_{1/2}}\\
\end{aligned}$}
& \vspace{-.7em}
Anti-inflammatory cytokines $C_A$ change according to:
\begin{itemize}[noitemsep,topsep=0pt,leftmargin=*]
\item production at constant rate $s_c$; decay at rate $\mu_c$
\item production related to phagocyte and tissue damage levels
\item stimulation by effector cells $|\vec{E}|$
\item \textcolor{black}{1 in denominator of second term has units of [$N^*$]}
\end{itemize}\baselineskip 0pt
\\\hline%
\raisebox{-3.2em}{
$\begin{aligned}
&R = f(k_{nn} N^* + k_{np} P + k_{nd} D,\ \textcolor{black}{C_A})\\
&f(x,\ C_A) = \frac{x}{1 + \left(\frac{C_A}{C_\infty}\right)^2}\\
&f_s(y) = \frac{y^6}{x_{dn}^6 + y^6}
\end{aligned}$}
& \vspace{-.8em}
\begin{itemize}[noitemsep,topsep=0pt,leftmargin=*]
\item $R$ is an aggregation of signals that trigger the innate immune response
\item $f(x,\ C_A)$ mediates the value of $x$ according to the level of anti-inflammation cytokines $C_A$
\item $f_s(y)$ was phenomenologically fit by Reynolds \textit{et al}.~in their original formulation \cite{ReynoldsErmentrout2006}
\end{itemize}\baselineskip 0pt
\\
\hline
\end{tabular}
\caption{{\bf Full equations of the integrated immune branch (IIB) model that govern the dynamics of the immunological state variables described in Table \ref{tab:BioComponents}}. A full list of parameter values and descriptions is given in Table~\ref{tab:Params}}.
\label{tab:model_eqns}
}
\end{table}

\begin{table}[t!]

\scriptsize
\begin{tabular}{| cl p{3.6cm} c| cl p{3.6cm} c|}
\hline
\rowcolor[HTML]{9B9B9B}
Parameter & Value & Description and \textcolor{black}{dimension} & Source & Parameter & Value & Description and \textcolor{black}{dimension} & Source\\ 
\hline
$k_{pg}$ & {\bf 0.6} & pathogen logistic growth rate; [$T^{-1}$]& \cite{ReynoldsErmentrout2006} &
    $P_\infty$ & {\bf 20} & pathogen logistic carrying \newline capacity; [$P$]  &\cite{ReynoldsErmentrout2006}\\
\rowcolor[HTML]{D9D9D9} 
$k_{pm}$ & {\bf 0.6} & pathogen clearance rate by \newline nonspecific response; [$T^{-1}$] & \cite{ReynoldsErmentrout2006}&
    $s_m$ & {\bf 0.005} & source rate of nonspecific \newline response; [$T^{-1}$] & \cite{ReynoldsErmentrout2006}\\
$\mu_m$ & {\bf 0.002} & decay rate of nonspecific \newline response; [$T^{-1}$] & \cite{ReynoldsErmentrout2006}&
    $k_{mp}$ & {\bf 0.01} & rate of nonspecific exhaustion per pathogen; [$P^{-1}T^{-1}$] &\cite{ReynoldsErmentrout2006}\\
\rowcolor[HTML]{D9D9D9} 
$k_{pn}$ & {\bf 1.8} & rate of pathogen clearance by \newline innate response; \newline [$(N^*)^{-1}T^{-1}$] & \cite{ReynoldsErmentrout2006}&
    $\gamma$ & 0.02 & binding \textcolor{black}{rate} between \newline pathogens and adaptive cells of the same type; [$C^{-1}T^{-1}$] & \cite{StrombergCarlson2006}\\
$p$ & 0.9 & proportion of effector cell \newline resources allocated to \newline pathogen clearance; \newline [nondim.]&& 
    $\Delta$ & 0.1 & rate of antigen presentation; [$T^{-1}$]  &\\
\rowcolor[HTML]{D9D9D9} 
$\beta$ & 0.01 & \textcolor{black}{efficacy} of $P_j^{(3)}$ depletion by \newline antigen presentation; \newline [nondim.] && 
    $\alpha$ & 0.1 & \textcolor{black}{efficacy} of adaptive cell \newline activation by antigen \newline presentation; [$CP^{-1}$]  & \cite{StrombergCarlson2006}\\
$\theta_N$ & 5 & naive cell creation rate; \newline [$CT^{-1}$] & &
    $R_0$ & 7200 & total naive and memory cell logistic carrying capacity; [$C$]& \\
\rowcolor[HTML]{D9D9D9} 
$\delta_N$ & 0.025 & naive cell homeostasis rate; \newline [$T^{-1}$] &&
    $\delta_M$ & 4e-5 & memory cell decay rate; \newline [$T^{-1}$] & \\
$f$ & 0.4 & proportion of memory and \newline naive cells that divide into \newline effector cells; [nondim.] &\cite{StrombergCarlson2006}& 
    $\delta_E$ & 0.05 & effector cell decay rate; [$T^{-1}$]  &\cite{StrombergCarlson2006}\\
\rowcolor[HTML]{D9D9D9} 
$s_{nr}$ & {\bf 0.08} & maximum phagocyte \newline recruitment rate; [$N^*T^{-1}$] & \cite{ReynoldsErmentrout2006}&
    $\mu_{nr}$ & {\bf 0.12} & phagocyte recruitment\newline half-saturation constant; \newline [$T^{-1}$] &\cite{ReynoldsErmentrout2006}\\
$\mu_n$ & {\bf 0.05} & phagocyte decay rate; [$T^{-1}$]& \cite{ReynoldsErmentrout2006}&
    $k_{dn}$ & {\bf 0.35} & rate of tissue damage due \newline to phagocytes; [$DT^{-1}$] &\cite{ReynoldsErmentrout2006} \\
\rowcolor[HTML]{D9D9D9} 
$\mu_d$ & {\bf 0.02} & tissue damage decay rate; \newline [$T^{-1}$] &\cite{ReynoldsErmentrout2006}& 
    $s_c$ & {\bf 0.0125} & source rate of \newline anti-inflammatory cytokines; [$C_AT^{-1}$]& \cite{ReynoldsErmentrout2006}\\
$k_{cn}$ & {\bf 0.04} & maximum activation of \newline anti-inflammatory cytokines \newline by phagocytes and tissue \newline damage [$C_AT^{-1}$]&\cite{ReynoldsErmentrout2006}&
    $k_{cnd}$ & {\bf 48} & conversion rate between \newline tissue damage and phagocyte \newline abundance; [$N^*D^{-1}$] & \cite{ReynoldsErmentrout2006}\\
\rowcolor[HTML]{D9D9D9} 
$\mu_c$ & {\bf 0.1} & anti-inflammatory cytokine \newline decay rate; [$T^{-1}$] &\cite{ReynoldsErmentrout2006}&
    $k_{nn}$ & {\bf 0.01} & conversion rate between \newline phagocyte abundance and \newline aggregate innate response $R$; [$(N^*)^{-1}T^{-1}$]&\cite{ReynoldsErmentrout2006} \\
$k_{np}$ & {\bf 0.1} &conversion rate between \newline pathogen abundance and \newline aggregate innate response $R$; [$P^{-1}T^{-1}$]&\cite{ReynoldsErmentrout2006}& 
    $k_{nd}$ & {\bf 0.02} &conversion rate between \newline tissue damage and aggregate innate response $R$; [$D^{-1}T^{-1}$] &\cite{ReynoldsErmentrout2006}\\
\rowcolor[HTML]{D9D9D9} 
$k_{ce}$ & 0.4 & maximum anti-inflammatory \newline cytokine production rate by \newline effector cells; [$C_AT^{-1}$] && 
    $E_{1/2}$ & 10 & half-saturation constant for\newline  cytokine production by \newline effector cells; [$C$]& \\
$C_{\infty}$ & {\bf 0.28} &scaling factor for\newline  anti-inflammatory cytokine \newline abundance; [$C_A$]&\cite{ReynoldsErmentrout2006}& 
    $x_{dn}$ & {\bf 0.06} &phenomenologically-inferred \newline half-saturation constant; \newline [$N^*$] & \cite{ReynoldsErmentrout2006}\\
\rowcolor[HTML]{D9D9D9} 
$S_{max}$ & {\bf 36} & number of pathogen shapes in \newline shape space; [nondim.] & \cite{StrombergCarlson2006} & &&&
    \\
\hline
\end{tabular}
\caption{{\bf Typical parameters of the immune model in Table \ref{tab:model_eqns}.}
The parameters listed in this table are used to generate Fig.~\ref{fig:multi_infec}, while the other figures are created with slightly modified parameters as detailed in the Supplementary Information.
\blue{
Most innate parameters were originally described in the Reynolds {\em et al.} model \cite{ReynoldsErmentrout2006}, while most adaptive parameters were originally described in the Stromberg and Carlson model \cite{StrombergCarlson2006}.
Parameter values that are the same as those used in the original models are bold-faced.
\textcolor{black}{Dimensions are given in square brackets, with [$T$] denoting time, other symbols denoting the concentrations of their corresponding immune variables, and [$C$] denoting concentrations of adaptive immune cells (i.e. naive cells $N_i$, memory cells $M_i$, or effector cells $E_i$).}}
\label{tab:Params}}
\end{table}

\subsection{Adaptive immune response}
The IIB model also incorporates the adaptive immune response (shown in the magenta box of Fig.~\ref{fig:schematic}), which is based on the shape-space adaptive immune model of Stromberg and Carlson \cite{StrombergCarlson2006}.
By assigning discrete ``shapes'' to pathogen epitopes and adaptive immune cells, this formulation allows pathogen-specific immune memory and immune responses to be developed \cite{PerelsonOster1979}.
Thus, an introduced pathogen $P_i$ of shape $i$ induces an adaptive response consisting of naive cells $N_i$, memory cells $M_i$, and effector cells $E_i$ that all specifically target the introduced pathogen.
There are $S_{max}$ available shapes.
In the original model of Stromberg and Carlson, pathogens $P_i$ of shape $i$ could interact with adaptive cells of some different shape $j$ with a lowered binding affinity,
but in this paper for computational efficiency we require that adaptive responses be activated by a pathogen of identical shape \textcolor{black}{(that is, cross-reactivity is not implemented in the IIB model)}.

The process of antigen presentation delays the activation of the adaptive immune response. 
\textcolor{black}{To model this phenomenon using ordinary differential equations (rather than with delay differential equations)}, the delay is hard-coded in the IIB model with the linear chain technique. This process is schematized in the orange box of Fig.~\ref{fig:schematic}, in which the populations $P_i^{\text{(1)}}$, $P_i^{\text{(2)}}$, and $P_i^{\text{(3)}}$ are intermediate states that represent different stages of antigen presentation \cite{macdonald1978lecture}.
Eventually, the presented antigen $P_i^{(3)}$ induces naive cells $N_i$ to divide into memory cells $M_i$ and effector cells $E_i$; and memory cells $M_i$ to divide into additional memory cells $M_i$ and effector cells $E_i$.
Once created, these pathogen-specific effector cells $E_i$ work to clear the pathogen $P_i$.


\subsection{Integrated immune branch (IIB) model}
The IIB model is described in full in Table~\ref{tab:model_eqns}, with descriptions and values of the parameters given in Table~\ref{tab:Params}.
Next, we emphasize the modifications that synthesized the two separate innate and adaptive immune models.

In the IIB model, the innate and adaptive components are linked in two ways.
First, the two compartments are implicitly linked through the pathogen population: a higher pathogen load $P_i$ not only activates more phagocytes $N^*$ in the innate compartment, but also leads to a higher rate of antigen presentation and subsequent activation of naive $N_i$, memory $M_i$, and effector $E_i$ cells.
Thus, depletion of pathogen by either response (by phagocytes $N^*$ or by effector cells $E_i$) affects both compartments.

Second, the two compartments are explicitly linked since effector cells $E_i$ can create anti-inflammatory cytokines $C_A$ that weaken the innate response. 
The suppression of inflammatory responses by effector cells has been observed experimentally \cite{guarda2009t, palm2007not, kim2007adaptive}.
In the IIB model, effector cells are allocated either to clear pathogens or to promote the production of anti-inflammatory cytokines, in proportion to $p$ and $1-p$, respectively (dashed line in Fig.~\ref{fig:schematic}).
The pathogen removal efficiency $p$ may be varied from 0 to 1, so that in the limit that effector cells are solely responsible for pathogen clearance ($p=1$) no additional anti-inflammatory cytokines are produced.
Therefore, the synthesized model is capable of quantitatively
comparing the two adaptive immune functions of pathogen clearance and inflammation attenuation.

The IIB model introduces homeostatic constraints that regulate the capacity of naive and memory cells. 
The rates of both homeostatic responses are dependent on the sum of the bulk naive cell population $|\vec{N}| \equiv \sum_i N_i$ and the bulk memory cell population $|\vec{M}| \equiv \sum_i M_i$. 
In the process of clearing a pathogen, memory cells accumulate.
Afterwards, to satisfy the homeostatic constraints, naive cells must become less abundant than they were before the infection.
In the Supplementary Information, we derive analytic approximations to the immune dynamics that are generated by these homeostatic relaxations; when the timescale of pathogen clearance is much shorter than the timescale of homeostatic relaxation, these expressions can be used as part of a dynamic programming approach to significantly speed up numerical simulations. 


\section{Results}

\begin{figure}[t!]
\begin{center}
\includegraphics[width=.7\textwidth]{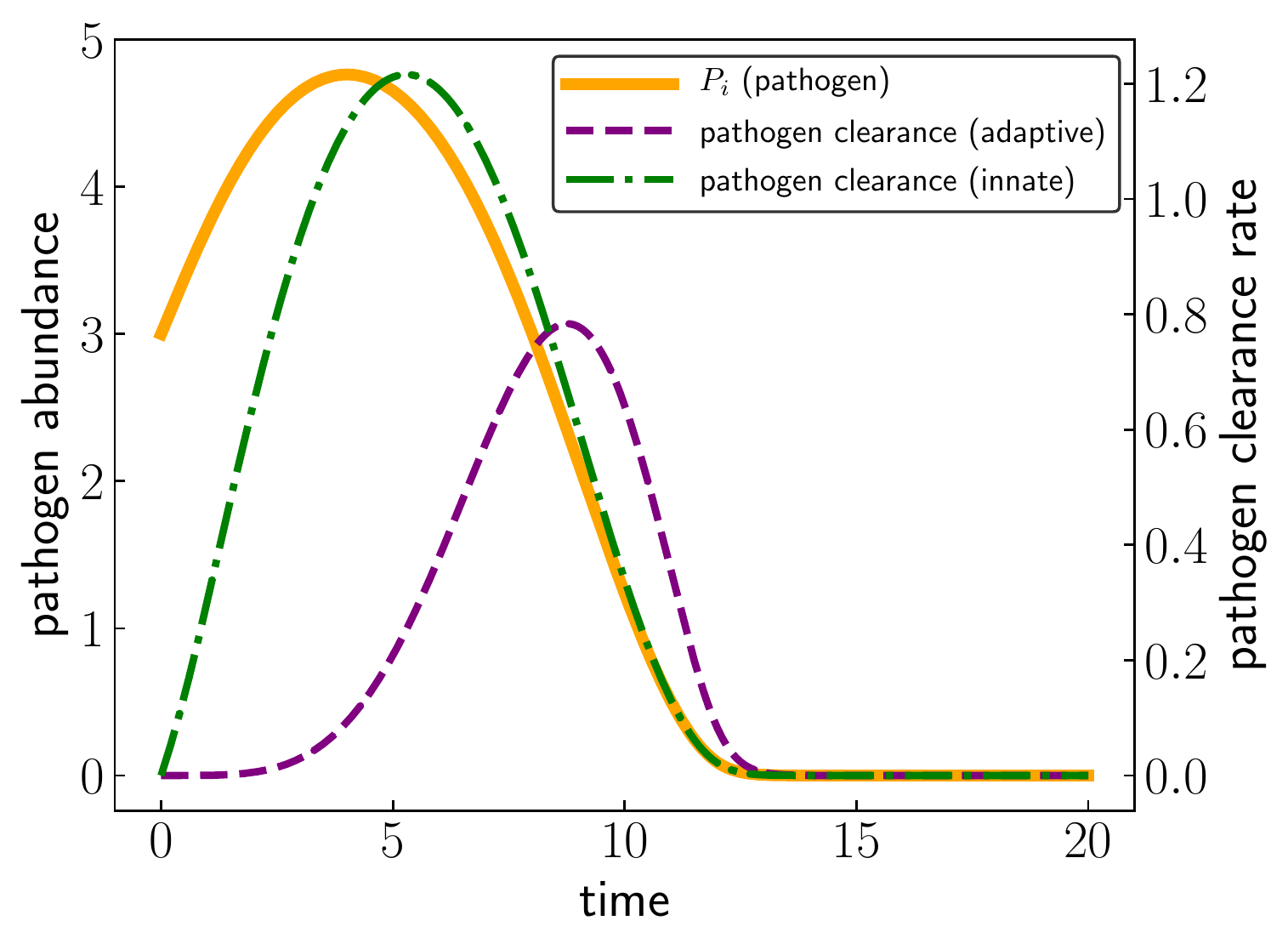}
\end{center}
\caption{{\bf Pathogen abundance (orange, solid) is regulated by the innate and adaptive immune responses in the IIB model.}
The clearance rates of the innate response (green dash-dotted, rate given by $k_{pn} f(N^*) P_i$) and the adaptive response (purple dashed, rate given by $p \gamma P_i E_i$), as described in the $\frac{\text{d}P_i}{\text{d}t}$ equation of Table~\ref{tab:model_eqns}, are plotted.
The innate response is activated immediately, while the adaptive response is delayed due to the antigen presentation process (encoded with the linear chain technique $P_i \to P_i^{(1)} \to P_i^{(2)} \to P_i^{(3)}$).
Ultimately the combined immune responses manage to clear the pathogen.
The parameters used to generate this figure are given in Table \ref{tab:Params} and in Table S1 of the Supplementary Information.
\label{fig:single_infection}}
\end{figure}

\subsection{Clearance of a single infection by the coupled immune response}
\begin{figure}[t!]
\begin{center}
\includegraphics[width=\textwidth]{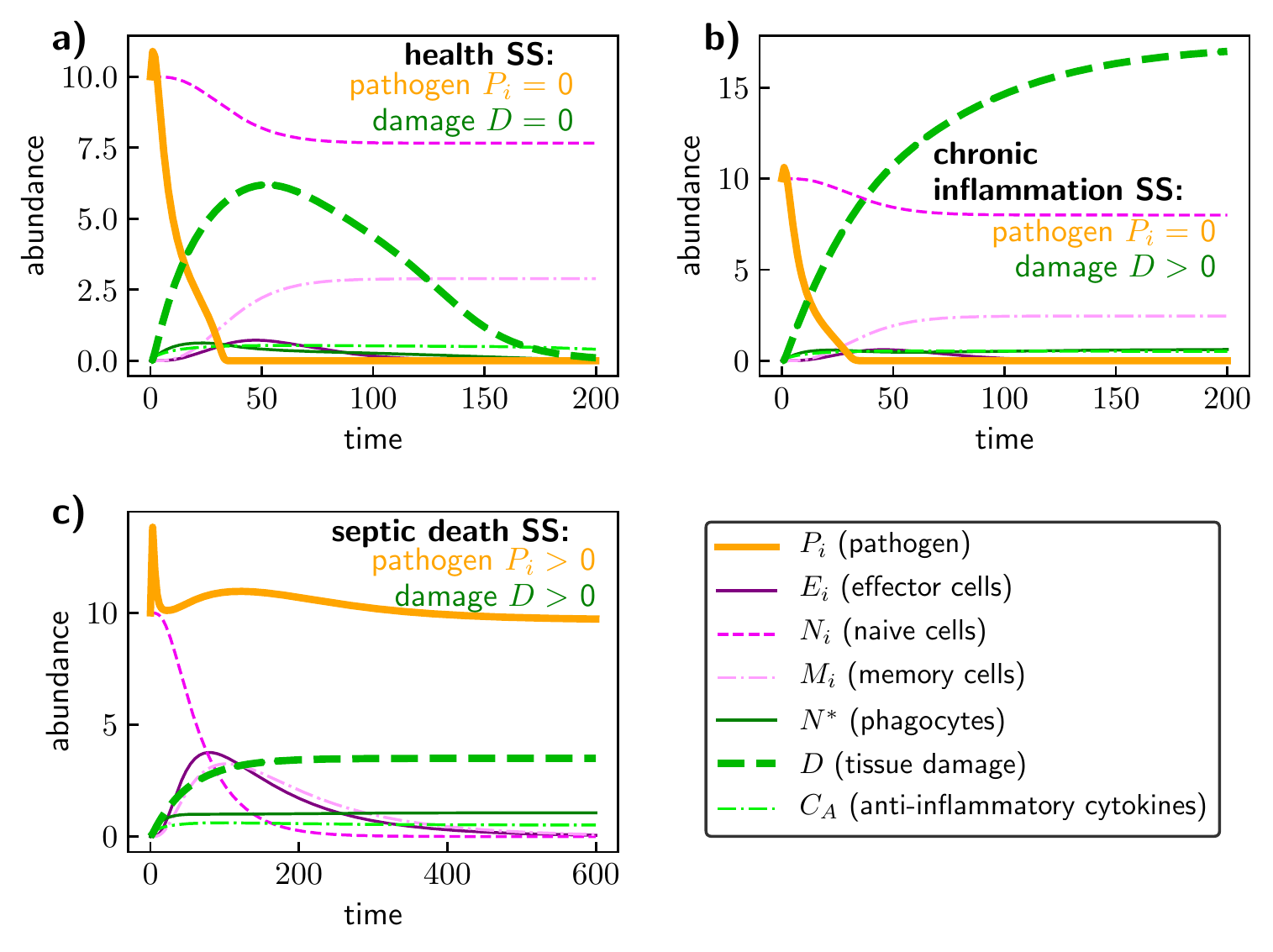}
\end{center}
\caption{{\bf The IIB immune model exhibits (a) health, (b) chronic inflammation, and (c) septic death steady states.}
Components of the adaptive response are plotted in purple, while components of the innate response are plotted in green.
(a) An inoculated pathogen activates phagocytes, which in turn induce tissue damage. 
Following antigen presentation, naive cells divide into memory cells and effector cells.
The phagocytes and effector cells jointly suppress the pathogen, which goes extinct, and the tissue damage gradually decays resulting in the health steady state.
(b) The innate and adaptive immune responses clear the pathogen, but in the process the innate response enters a positive feedback loop between phagocyte recruitment and tissue damage leading to persistent tissue damage and phagocyte activation, called the chronic inflammation steady state.
(c) The innate and adaptive immune responses do not clear the pathogen, leading to the septic death steady state characterized by the presence of pathogen and tissue damage.
The chronic inflammation steady state was obtained with smaller innate clearance rate $k_{pm}$ and smaller tissue damage decay rate $\mu_d$ than were used to obtain the health steady state.
The septic death steady state was obtained with a larger proportion of cognate cells that divide into effector cells $f$ and a larger pathogen growth rate $k_{pg}$ than were used to obtain the health steady state.
Explicit values of the parameters used for each panel are given in Table \ref{tab:Params} and in Table S1 of the Supplementary Information.
\label{fig:steady_states}}
\end{figure}

In the IIB model, once a pathogen is introduced it initiates a cascade of immune responses in the innate and adaptive compartments, and these responses combat the logistic growth of the pathogen and attempt to drive it to extinction.
Fig.~\ref{fig:single_infection} depicts a representative pathogen encounter and clearance, and plots the pathogen abundance (orange) as well as the pathogen clearance rates due to the adaptive effector cell (purple) and innate phagocyte (green) responses given by the quantities $p \gamma P_i E_i$ and $k_{pn} f(N^*) P_i$, respectively, as given in Table \ref{tab:model_eqns}. 
Note that the adaptive response is specific to the pathogen shape, while the innate response is nonspecific.
For clarity only the lumped contributions of the innate and adaptive compartments to pathogen clearance are plotted in Fig.~\ref{fig:single_infection}. 
The populations of every immunological variable are plotted for several scenarios in Fig.~\ref{fig:steady_states}.

In Fig.~\ref{fig:single_infection} the logistic growth of the pathogen drives the initial pathogen spike, which is mildly tempered by a non-local immune response as described by Reynolds \textit{et al.} \cite{ReynoldsErmentrout2006}.
Phagocytes are immediately recruited and attack the pathogen, leading to the increase in innate pathogen clearance (green).
Simultaneously, the collateral tissue damage inflicted by phagocytes causes the production of anti-inflammatory cytokines, which suppress further phagocyte recruitment and tissue damage.
Anti-inflammatory cytokines in conjunction with a decreasing pathogen population cause the decrease in innate pathogen clearance.
Once the pathogen is presented to the adaptive immune branch--- a delay that is hard-coded in the IIB model with the auxiliary immunological variables $P_i^{(1)}$, $P_i^{(2)}$, and $P_i^{(3)}$--- the naive and memory cells specific to the presented antigen divide into effector cells.
These effector cells subsequently contribute to the increase in adaptive pathogen clearance (purple).
Ultimately, with the given parameter values (provided in the Supplementary Information) the innate and adaptive responses overpower the pathogen and drive it to extinction.
In the process, memory cells specific to this pathogen shape proliferate and provide future protection in case the same pathogen is faced again in the future, since the higher initial abundance of pathogen-specific memory cells will result in a more immediate adaptive response. 

\begin{figure}[t!]
\begin{center}
\includegraphics[width=\textwidth]{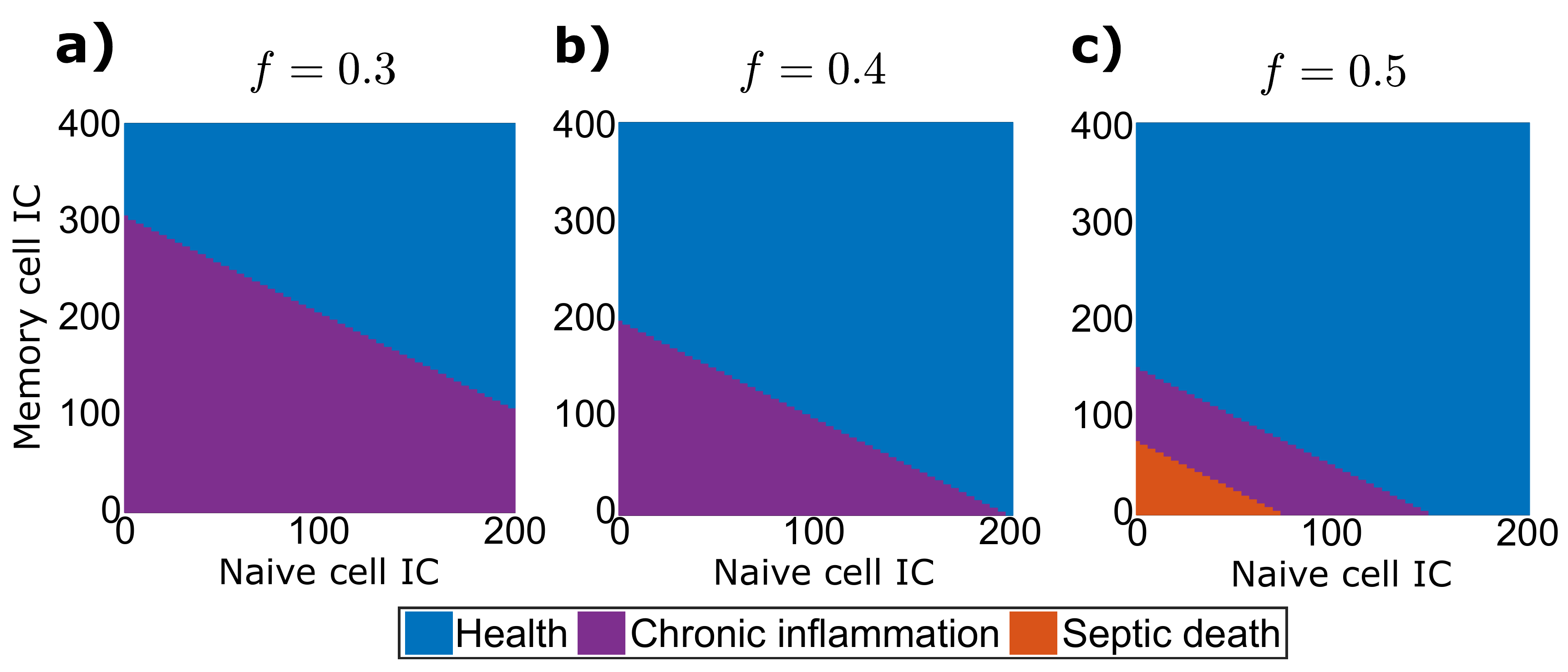}
\end{center}
\caption{{\bf Phase diagram of immunological steady states as a function of naive cell and memory cell initial conditions (ICs).}
In IIB model, the immune system can reach health, chronic inflammation, or septic death steady state following a pathogen encounter, depending on the level of cognate naive and memory cells when the pathogen is introduced.
\textcolor{black}{The proportion $f$ of cognate cells that divide into effector cells significantly influences the steady-state phase diagram; in particular, septic death (red zone) can only occur for $f\geq0.5$.}
Results are calculated in the absence of homeostatic response ($\theta_N=\delta_N=\delta_M=0$ in Table 2)  during single infection events and initial pathogen level is $P_i=1$. Other parameters for generating the phase diagrams are as stated in Table 3. Except in current figure, $f=0.4$ is always used in this study.
\label{fig:PhaseDiagram}}
\end{figure}
\subsection{Steady state analysis of the IIB model}
The IIB model described in Table \ref{tab:model_eqns} exhibits steady states when the time derivatives of all the populations vanish.
This coupled model inherits many of the steady state characteristics of the constituent innate and adaptive immune models.
In particular, this model exhibits steady states of (a) \textit{health}, characterized by vanishing pathogen and immune response; (b) \textit{chronic inflammation}, in which the pathogen clears but the innate immune response is sustained in a positive feedback loop; and (c) \textit{septic death}, characterized by the chronic presence of pathogen and activated immune responses. 
Realizations of these three steady states are displayed in Fig.~\ref{fig:steady_states}, and the parameter sets used to generate each outcome are provided in Table \ref{tab:Params} and in Table S1 of the Supplementary Information.
\textcolor{black}{The steady states attained by the IIB model are sensitive to parameter values, and in Fig.~\ref{fig:steady_states} different steady states were attained by varying the innate cell clearance rate $k_{pm}$, the tissue damage decay rate $\mu_d$, the proportion $f$ of cells that divide into effector cells, and the pathogen growth rate $k_{pg}$.}
In Fig.~\ref{fig:PhaseDiagram}, phase diagrams demonstrate how different parameter regimes and initial conditions lead to different steady states.
To observe the presence of these steady states mathematically, note that a quantity $P_i$ may be factored out of the pathogen dynamics $\frac{\text{d}P_i}{\text{d}t}$, from which it is clear that at steady state the pathogen population $\bar{P}_i$ must either be 0 or some nonzero quantity that satisfies $\frac{1}{P_i}\frac{\text{d}P_i}{\text{d}t}=0$. The three steady states immediately follow from the implications of choosing $\bar{P}_i$ to be zero or non-zero.

\textit{(a) Health steady state---}
If $\bar{P}_i=0$ at equilibrium, then the intermediate pathogen states must vanish as well ($\bar{P}_i^{(1)} = \bar{P}_i^{(2)} =\bar{P}_i^{(3)} = 0$).
In the absence of presented antigen, the memory cells $M_i$ decay with timescale $1/\delta_M$.
When this timescale is slow relative to the homeostatic dynamics of the naive cells (whose dynamics are of timescale $1/\delta_N$), the naive cells $N_i$ tend towards their homeostatic equilibrium $\bar{N}_i$ as described in Eq.~(S11) in the Supplementary Information.
Lastly, all effector cells $E_i$ decay with timescale $1/\delta_E$, which is assumed to be fast compared to the naive and memory cell dynamics.
Thus, in the absence of pathogen, the adaptive immune response turns off and becomes dormant.
In a steady state with $\bar{P}_i=0$ the innate immune response can either be ``inactive'' or it can be ``active,'' which lead to the steady states of \textit{health} or \textit{chronic inflammation}, respectively.
When the innate immune response is inactive, the activated phagocytes $N^*$  and tissue damage $D$ are both zero.
This implies that the aggregate innate response $R$ is zero as well.
Lastly, in the absence of the innate response and effector cells, anti-inflammatory cytokines equilibrate to a constant level. 
It is straightforward to check that setting $\bar{P}_i = \bar{N}^* = \bar{D} = 0$ and $\bar{C}_A = s_c/\mu_c$ leads to a steady state in this model.

This health steady state is demonstrated numerically in Fig.~\ref{fig:steady_states}a. 
In this figure, an initial pathogen response activates innate (green) and adaptive (purple) responses that eventually vanish.
The naive and memory cells change over the course of the adaptive immune response, and then remain at constant steady state values.

\textit{(b) Chronic inflammation steady state---}
When $\bar{P}_i=0$ but the innate immune response is active, the model reaches the chronic inflammation steady state.
This steady state is inherited from the Reynolds \textit{et al}.~innate immune model \cite{ReynoldsErmentrout2006}, and occurs due to a positive feedback loop between tissue damage $D$ and phagocytes $N^*$ (as can be schematically understood based on Fig.~\ref{fig:schematic}).
In particular, there exist equilibrium quantities $\bar{N}^*$ and $\bar{D}$ that precisely balance the activation of phagocytes and accumulation of tissue damage with their respective decays $\mu_n \bar{N}^*$ and $\mu_d \bar{D}$. 
This chronic inflammation steady state is shown in Fig.~\ref{fig:steady_states}b: the pathogen is cleared and effector cells dissipate, but the innate response is perpetually sustained.

\textit{(c) Septic death steady state---}
Lastly, the steady state in which the pathogen population is sustained is called septic death.
For a steady state with nonzero pathogen $\bar{P}_i$, the values of the intermediate pathogen states $\bar{P}_i^{(1)}$, $\bar{P}_i^{(2)}$, and $\bar{P}_i^{(3)}$ will be nonzero as well.
Subsequently, the presented pathogen $\bar{P}_i^{(3)}$ sustains the activation of naive, memory, and effector cells.
In the innate compartment, the nonzero pathogen presence implies a nonzero aggregate innate response $\bar{R}$, which implies a nonzero equilibrium population of phagocytes $\bar{N}^*$, which in turn implies a nonzero equilibrium population of tissue damage $\bar{D}$.
Therefore, the septic death steady state is characterized by activity in both the innate and the adaptive immune compartments.
Naive and memory cells of the adaptive compartment continue to predominantly divide into effector cells; eventually all adaptive cells are exhausted and vanish while failing to clear the pathogen. 
This steady state is depicted in Fig.~\ref{fig:steady_states}c.

The rest of this paper uses a parameter regime that does not exhibit the septic death steady state\blue{: in particular, simulations of this paper set $f$ (the proportion of cognate cells that divide into effector cells) equal to $0.4$, which causes memory cells to accumulate until they are able to produce enough effector cells to suppress the pathogen.
When $f$ is larger than $0.5$ memory cells deplete over the course of an immune response, which can lead to septic death.
A phase diagram of steady state behaviors at different values of $f$ for different naive and memory cell initial conditions is plotted in Fig.~\ref{fig:PhaseDiagram}}.
\textcolor{black}{In particular, septic death is only reachable when $f \geq 0.5$: if $f < 0.5$ the number of memory cells will strictly increase over time, eventually leading to a sufficiently strong adaptive immune response capable of clearing any pathogen (and thus prohibiting septic death). 
}
\blue{In what follows,} we focus on the transition from health to chronic inflammation, a process phenomenologically similar to inflammaging.

\begin{figure}[t!]
\begin{center}
\includegraphics[width=.8\textwidth]{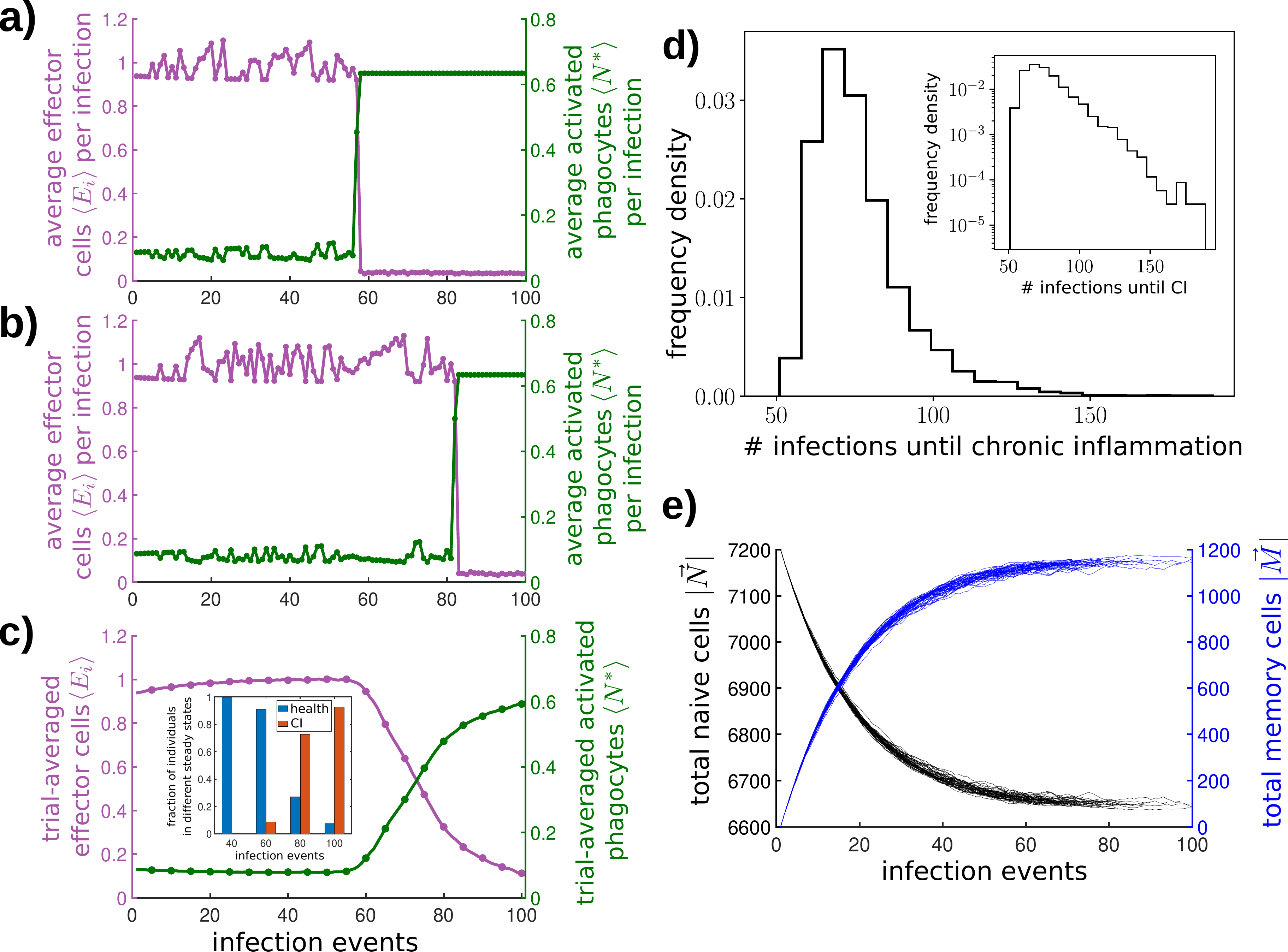}
\end{center}
\caption{{\bf Timing of the transition to chronic inflammation (CI) is highly variable and depends on previous pathogen encounters.} 
(a, b) Activity of the innate (green) and adaptive (purple) immune responses over the course of 100 infection events for two different infection sequences drawn from the same statistical distribution.
\textcolor{black}{These regularly-spaced infection events are used to measure the age of the immune system.}
As a proxy for these responses, the average number of effector cells $\langle E\rangle\equiv\frac{1}{\Delta T}\int_{t_\ell}^{t_{\ell+1}} |\vec E(t)| \ \text dt$ and phagocytes $\langle N^*\rangle \equiv \frac{1}{\Delta T}\int_{t_\ell}^{t_{\ell+1}} N^*(t) \ \text dt$ for each infection event $\ell$ are plotted.
The sharp transitions indicate the onset of the chronic inflammation state, and occur at the $57$th and $82$nd infection events, respectively.
(c) An ensemble average of the innate and adaptive immune responses over 1000 infection sequences (each drawn from the same pathogen distribution) smooths the variability in transition timing, though the distribution of immune responses is bimodal (c, inset). CI: chronic inflammation
(d) The distribution of transition times to chronic inflammation is concentrated at earlier times (on average after 76 infections).
(e) The accumulation of memory cells (blue) and depletion of naive cells (black) drives immune fragility and vulnerability to new pathogen shapes (50 infection sequences shown).
These figures are generated with the parameters given in Table \ref{tab:Params}, and with randomly generated pathogen sequences as described in Eq.~\ref{eq:Ps}.
\label{fig:multi_infec}}
\end{figure}

\subsection{The onset of chronic inflammation results \textcolor{black}{from a fragility induced by} a lifetime of infections}
Next we consider the immunological consequences that result from encountering a sequence of infection events.
Infection sequences are composed of discrete infection events, and each infection event consists of the time course following the encounter with a particular pathogen shape
until a steady state of the system (i.e. health, chronic inflammation, or septic death) is reached.
For each infection event an immune response is generated by simulating the IIB model, given in Table \ref{tab:model_eqns}.
When pathogen encounters are evenly spaced in time, the number of infection events acts as a measure of age. 
This infection sequence encodes a lifetime of infection events, is different for different individuals, and serves as a vehicle with which to explore the variable immunological outcomes experienced by different individuals over their lifetimes.



\begin{figure}[t!]
\begin{center}
\includegraphics[width=0.8\textwidth]{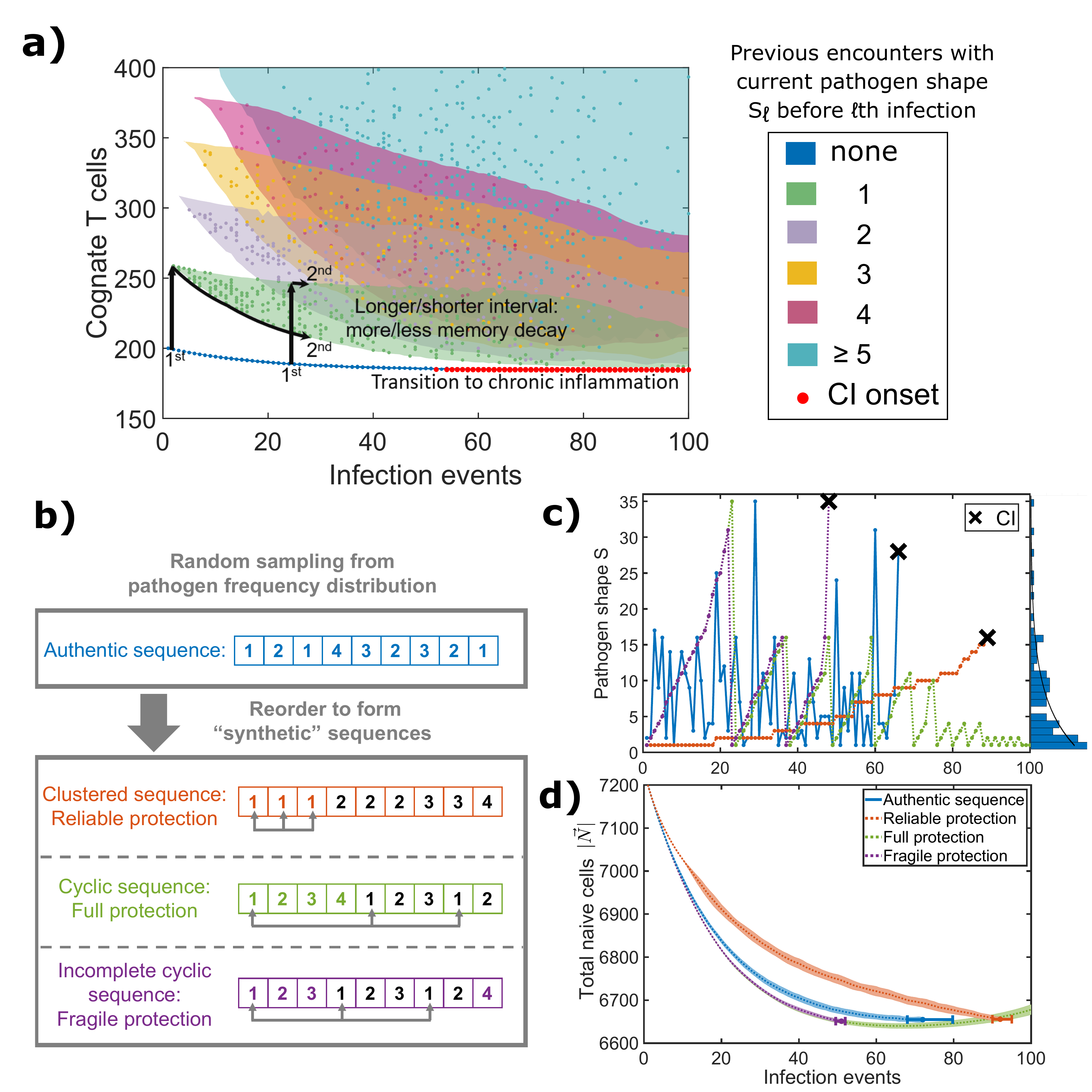}
\end{center}
\caption{{\bf \textcolor{black}{Aging-induced} transition to chronic inflammation (CI) is driven by depletion of naive cells and lack of protection from memory cells.}
\textcolor{black}{The number of encountered infections is used as a proxy for the age of the immune system.}
\blue{(a) The number of cognate T cells specific to a novel pathogen shape (equal to the sum of naive and memory cells) is the key indicator for whether an infection event will trigger the chronic inflammation steady state.
Here, cognate T cell counts specific to an encountered pathogen $P_\ell$ are plotted for each infection event $\ell$ across 20 infection sequences sampled from Eq.~(\ref{eq:Ps}).
The color of each point indicates the number of times that the encountered pathogen $P_\ell$ has been previously encountered.
The colored bands are generated from 1000 infection sequences, and envelope the observed cognate cell counts. 
%
The large red circles in the lower-right corner mark the infections events that trigger chronic inflammation across all 1000 infection sequences, which occur when a novel pathogen is encountered after naive cells have been depleted below some threshold.
A shorter time interval between pathogen encounters of the same shape results in less memory cell decay and hence more cognate T cells, and this effect causes the shape of the colored bands.}
(b) We consider three synthetic reorderings of each ``authentic'' randomly generated pathogen sequence: the clustered sequence orders pathogens according to their prevalence; the cyclic sequence orders them to ensure immediate exposure to all pathogen types; 
and the incomplete cyclic sequence induces fragility by quickly depleting naive cells and then introducing a novel pathogen.
(c) The authentic sequence and three synthetic sequences transition to chronic inflammation (CI) at different times (black crosses).
The pathogen shape distribution for this infection history (right histogram) is drawn from the
theoretical shape distribution (black line overlaid) given by Eq.~(\ref{eq:Ps}). 
(d) The naive cell pool is depleted at different rates depending on how infection events are ordered.
Naive cell counts and their variation across the 50 authentic sequences considered in panel (a) are shown for the three synthetic sequences.
Error bars for the timing of chronic inflammation are 50\% confidence intervals.
\label{fig:ASD_mechanism}}
\end{figure}

More concretely, the IIB model is simulated for $n_{tot}$ infections that are $\Delta T = 1000$ time units apart. 
The time $\Delta T$ is chosen to be sufficiently large so that the system reaches a steady state between infection events (i.e., infection events are well separated in time).
The shape space in the IIB model is discrete, consisting of $S_{max}=36$ available pathogen shapes. 
For each infection event, the probability $p_i$ that a pathogen of shape $i$ is encountered is given by 
\begin{equation} \label{eq:Ps}
p_i=\zeta e^{-i/\xi}, \quad i=1,\ 2, \ldots,\ S_{max},
\end{equation} 
where $\xi=20/3$ and $\zeta$ satisfies $\sum_{i=1}^{S_{max}} p_i=1$.
This distribution allows for some pathogen shapes to be more common than others, 
and is similar to the one originally used by Stromberg and Carlson \cite{StrombergCarlson2006}.    
A lifetime of infections is explicitly encoded in the infection sequence $\{S_\ell\}$, where each $S_\ell$ is the pathogen shape encountered for the $\ell$th infection event.
Then, for each infection event $\ell=1,\ 2,\ldots,\ n_{tot}$, a single unit of pathogen $P_{S_\ell}$ is added to the system at time $t_\ell = 1000(\ell-1)$.
In the simulations used to create Figs.~\ref{fig:multi_infec}, \ref{fig:ASD_mechanism}, and \ref{fig:pVary}, the system is initialized with zero memory cells $M_i(0)=0$ and a uniform distribution of naive cells  $N_i(0) = 200$ across all possible pathogen shapes $i=1,\ 2,\ldots,\ S_{max}$.

There are four important timescales in the IIB model: the time $\tau_{\textrm{infec}}$ required for pathogen clearance, the interval $\Delta T$ between infection events, and the timescales of naive and memory cell homeostasis control.
Pathogen clearance is the fastest process, during which the homeostasis control still has little effect, \textcolor{black}{and its timescale is on the order of days \cite{ReynoldsErmentrout2006}}. The timescales of naive and memory cell homeostasis are characterized by the reciprocal of their decay rates, given by $1/\delta_N$ and $1/\delta_M$, respectively. \textcolor{black}{Experimental data suggest their orders as months \cite{de2013quantifying} and decades or longer \cite{choo2010homeostatic}, respectively.}
Due to the longevity of immune memory, the interval between infections $\Delta T$ was chosen to be much shorter than the timescale of memory decay.
Additionally, as in Stromberg and Carlson naive cells are assumed to regenerate and equilibrate quickly relative to the rate at which infection events occur
\cite{StrombergCarlson2006}.
Thus, between infection events the homeostatic naive cell population depends on a slowly-decaying memory population. 
More explicitly, the four timescales in IIB model are chosen to satisfy $\tau_{\textrm{infec}} < 1/\delta_N < \Delta T \ll 1/\delta_M$.

\textcolor{black}{As the immune system ages (i.e., over the course of an infection sequence)}, early infection events (e.g. before the 50th infection event) are successfully cleared by the immune system, and the system returns to the health steady state.
However, for later infection events the system fails to recover and instead transitions to the chronic inflammation steady state, where it remains thereafter.
This age-driven, history-dependent transition to chronic inflammation is qualitatively similar to ``inflammaging.'' 

Depending on the details of the infection sequence, the timing of the onset of chronic inflammation is highly variable.
Two instances of the transition to chronic inflammation, with different sequences of pathogen encounters generated from the same statistical distribution of pathogen frequencies, are displayed in Fig.~\ref{fig:multi_infec}a and Fig.~\ref{fig:multi_infec}b.
The strengths of the adaptive (purple) and innate (green) responses during each infection event are plotted in Fig.~\ref{fig:multi_infec}a and Fig.~\ref{fig:multi_infec}b, quantified by the average number of effector cells $\langle E\rangle$ and average number of phagocytes $\langle N^*\rangle$ over the course of each infection, respectively.
For the $\ell$th infection,
\begin{equation}
\begin{aligned}
\langle E\rangle &\equiv \frac{1}{\Delta T}\int_{t_\ell}^{t_{\ell+1}} |\vec E(t)| \ \text dt, \quad \text{and}\\
\langle N^*\rangle &\equiv \frac{1}{\Delta T}\int_{t_\ell}^{t_{\ell+1}} N^*(t) \ \text dt,
\end{aligned}
\label{eq:EN_mean}
\end{equation}
where $|\vec E(t)|\equiv\sum_{i=1}^{S_{max}}E_i(t)$, and $t_\ell$ is the starting time of the $\ell$th infection event.
Once the system attains the chronic inflammation steady state the heightened inflammatory response will \textcolor{black}{rapidly clear pathogens in any future infection, thus} limiting the activity of the adaptive response. 
Accordingly, this leads to the sharp transition behavior of the two trajectories observed in Fig.~\ref{fig:multi_infec}a and \ref{fig:multi_infec}b. 
Therefore, the onset of chronic inflammation causes the innate response to strengthen while the adaptive response weakens.

When averaged over an ensemble of infection sequences, variability in the timing of the sharp transition from health to chronic inflammation smooths into the crossing displayed in Fig.~\ref{fig:multi_infec}c (though the actual distribution of effector cells and phagocytes across infection sequences is bimodal, as seen in the inset of Fig.~\ref{fig:multi_infec}c).
\blue{
This crossing behavior is consistent with a longitudinal study of Swedish people, in which
middle-aged people exhibited steady lymphocyte and neutrophil counts over the three-year span of the study, while older people ($>$85 years of age) exhibited significantly increased neutrophil counts and significantly decreased lymphocyte counts over the same span \cite{wikby1994age}.}
In addition, people with an immune risk phenotype (IRP) (which often precedes immune decline and death) commonly possess a weakened adaptive immune repertoire and experience inflammaging \cite{wikby2006immune}.
Taking $\langle E \rangle$ and $\langle N^* \rangle$ as proxies for adaptive and innate immune function, the age-dependent transition in the IIB model resembles the shift in immune function experienced by people with an IRP.

The number of infection events that are encountered before chronic inflammation is reached is plotted as a distribution across 10,000 randomly-generated infection sequences in Fig.~\ref{fig:multi_infec}d.
The distribution in Fig.~\ref{fig:multi_infec}d decays approximately exponentially for large numbers of infection events.
Different choices of pathogen shape distributions result in different distributions of transition times that are qualitatively similar but in general not exponentially distributed, as demonstrated in Fig.~S1.
Thus, the sequence of infection events is a significant driver of variability in the timing of the transition to chronic inflammation.

To identify the relationship between an infection sequence and the transition from health to chronic inflammation, we examine the effect of accumulated pathogen exposures on the bulk immune state variables $|\vv{M}|\equiv\sum_{i=1}^{S_{max}}M_i$ and $|\vv{N}|\equiv\sum_{i=1}^{S_{max}}N_i$; see Fig.~\ref{fig:multi_infec}e.
Over the course of an infection sequence, memory cells are generated in response to new pathogen encounters at a faster rate than their decay.
When fewer novel pathogens are encountered, memory accumulation slows, since previously encountered pathogens are cleared more quickly and adaptive immune cells are stimulated for a shorter amount of time.
As the memory cell repertoire grows, the naive cell population shrinks according to the homeostatic constraints.
\textcolor{black}{Eventually, the immune system develops a fragility} to novel pathogens due to the depleted naive cell population \textcolor{black}{that results from aging};
once this fragility is developed, chronic inflammation will be triggered when any novel pathogen is encountered. 

\subsection{The transition to chronic inflammation is influenced by previous pathogen encounters}
The disparate immunological outcomes of different individuals demonstrated in Fig.~\ref{fig:multi_infec} are necessarily determined by the difference in their infection sequences, since the model equations in Table \ref{tab:model_eqns} are otherwise deterministic.
Until the transition to chronic inflammation, the system always returns to the health steady state.
In the health steady state most immune variables assume values that are independent of previous pathogen encounters; only the naive and memory cell populations occupy values that are potentially different after each infection event.
Therefore different infection sequences lead to differences in memory and naive cell populations, which in turn are directly responsible for the transition to chronic inflammation.

In Fig.~\ref{fig:ASD_mechanism}a, the number of cognate T cells (the sum of pathogen-specific naive and memory cells $N_{S_\ell} + M_{S_\ell})$ at the beginning of each infection event $\ell$ are plotted for 1000 infection sequences, each consisting of 100 infection events sampled from the pathogen shape distribution Eq.~(\ref{eq:Ps}).
{\color{black} 
Due to the accumulation of immune memory, the number of cognate T cells specific to a pathogen shape will be greater if that pathogen has been previously encountered.
To demonstrate this, the colored points in Fig.~\ref{fig:ASD_mechanism}a encode the number of times that a pathogen shape $S_\ell$ was encountered in the $\ell - 1$ previous infection events. 
A shorter time interval between infections of the same pathogen shape leads to less memory decay and more cognate T cells specific to that pathogen, as shown in the black trajectories in Fig.~\ref{fig:ASD_mechanism}a.
The average time before a pathogen shape is re-encountered is randomly distributed, which causes the variability of each color band (computed from 1000 simulated infection sequences).
The number of cognate cells specific to novel pathogens is plotted in blue: in this case no pathogen-specific memory cells exist, and so the cognate cell and naive cell counts are the same.
Therefore, the decline in the dark blue dots demonstrates the gradual decay of naive cell counts over the course of an infection sequence \textcolor{black}{(i.e., as the immune system ages)}.
}

The infection events that trigger the transition to chronic inflammation are indicated by the red circles in Fig.~\ref{fig:ASD_mechanism}a.
These transitions always occur (i) in response to a novel pathogen shape (red circles are laid on top of the blue band), and (ii) when the number of cognate naive cells falls below a threshold (approximately $180$ in Fig.~\ref{fig:ASD_mechanism}a).
When both conditions are met, the adaptive immune response is low in magnitude and unable to produce sufficient anti-inflammatory cytokines to suppress the innate immune branch.
This weakened adaptive immune response, itself a function of the infection sequence, is the principal driver of chronic inflammation in the IIB model.
\blue{Accordingly, the IIB model exhibits a ``robust yet fragile" behavior \cite{StrombergCarlson2006, carlson2000highly}: it is robust to frequently-encountered pathogens, yet fragile to novel pathogens.} 

\subsection{Manipulating immune system fragility via synthetic infection sequences}

\blue{To probe the variability in the timing of the transition to chronic inflammation \textcolor{black}{as the immune system ages}, we examine three synthetic infection sequences that are reorderings of
an ``authentic'' infection sequence sampled from Eq.~(\ref{eq:Ps}).
These sequences, showcased schematically in Fig.~\ref{fig:ASD_mechanism}b and detailed in the text below,  deliberately structure the order of pathogen encounters to induce different levels of fragility towards novel pathogen shapes.
The synthetic and authentic infection sequences affect the rate of memory cell accumulation and naive cell loss, and in turn, alter the timing of the onset of chronic inflammation.}
An example authentic sequence, along with its corresponding synthetic sequences, is illustrated in Fig.~\ref{fig:ASD_mechanism}c.
In Fig.~\ref{fig:ASD_mechanism}d naive cell statistics of 50 authentic infection sequences are compared with the statistics generated by their synthetic counterparts.

In the {\em clustered} synthetic ordering (Fig.~\ref{fig:ASD_mechanism}b-d, orange), the infection sequence is ordered so that the most common pathogens are encountered first and the rarest pathogens are encountered last.
In this case, pathogens are immediately reencountered so memory cells do not significantly decay between infections, and the accumulated immune memory causes an accelerated immune response that generates fewer memory cells.
Therefore, this reordering is a {\em lower bound} for the naive cell loss rate.
Indeed, the clustered sequence leads to the slowest loss of naive cells among the authentic and synthetic sequences in Fig.~\ref{fig:ASD_mechanism}d. 
The clustered sequence provides reliable protection that delays the transition to chronic inflammation until infection 89, compared with the authentic sequence that transitions after infection 66, as shown in  Fig.~\ref{fig:ASD_mechanism}c.

In the {\em cyclic} synthetic ordering (Fig.~\ref{fig:ASD_mechanism}b-d, yellow) infections are ordered so that all available pathogen types are encountered as early as possible: in this ordering each pathogen type is encountered once before any pathogen is encountered for the second time, then each pathogen type is encountered twice before any pathogen is encountered for the third time, and so on.
Constantly encountering new pathogen types drives the accumulation of memory cells and in turn naive cell loss at an accelerated rate.
Thus the cyclic sequence yields an {\em upper bound} for the naive cell loss rate, as in Fig.~\ref{fig:ASD_mechanism}d.  
At the same time, since this synthetic sequence is structured to front-load every pathogen type that can be encountered early in the infection history, the generated memory cells eventually provide full protection against each pathogen type, and the chronic inflammation state never occurs, as in Fig.~\ref{fig:ASD_mechanism}c.
Thus, broad exposure to pathogens early in an individual's infection life history can provide adaptive-mediated protection from chronic inflammation in the IIB model.

Lastly, the {\em incomplete cyclic} synthetic ordering (Fig.~\ref{fig:ASD_mechanism}b-d, purple) is similar in construction to the cyclic ordering, except that one rare pathogen is intentionally omitted from the initial pathogen cycles. 
Then, this pathogen is presented at a later time to trigger chronic inflammation.
The incomplete cyclic ordering induces a fragile immune response: naive cells deplete nearly as quickly as for the cyclic ordering, and incomplete immune memory coverage causes vulnerability to novel pathogens.
Thus, the onset of chronic inflammation is accelerated in the incomplete cyclic ordering, with the onset occurring during the 48th infection in  Fig.~\ref{fig:ASD_mechanism}c.

The clustered and cyclic synthetic immune histories demonstrate how some pathogen sequences can delay the onset of chronic inflammation, either by prolonging the abundance of naive cells or by quickly acquiring full immune memory coverage across all pathogen shapes.  
In contrast, the incomplete cyclic sequence demonstrates how pathogen sequences can induce immune fragility, by quickly depleting naive cells while remaining vulnerable to novel pathogens\blue{; alternatively, the incomplete cyclic sequence shows how the introduction of a new pathogen species, either through mutation or migration to a new environment, can break existing memory coverage and lead to immune fragility.
}

\begin{figure}[t!]
\begin{center}
\includegraphics[width=.6\textwidth]{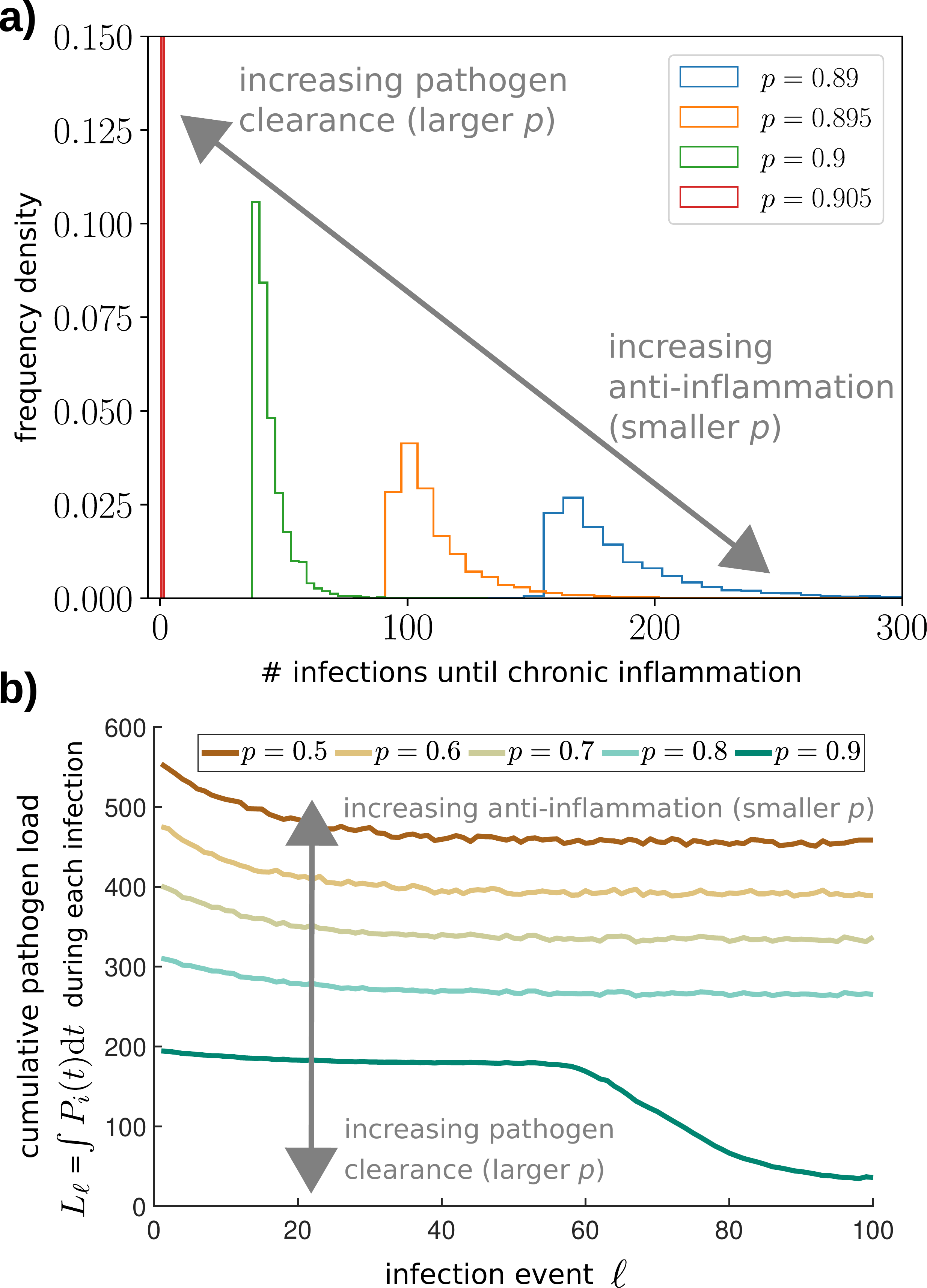}
\end{center}
\caption{{\bf Effector cells 
are subject to a trade-off between clearing pathogens and suppressing inflammation \textcolor{black}{as the immune system ages}.} 
(a) The onset of chronic inflammation (histograms as in Fig.~\ref{fig:ASD_mechanism}d) is delayed for lower values of $p$, i.e. when the anti-inflammatory role of effector cells is increased.
(b) The cumulative pathogen load $L_{\ell}$ over the course of each infection event (averaged over 1000 infection sequences) is larger for smaller values of $p$.
The drop in $L_{\ell}$ after the 60th infection event for $p=0.9$ is caused by the onset of chronic inflammation, which compensates for the overspecialized adaptive immune repertoire.
To generate the statistics in panel (a), the homeostatic parameters $\delta_N$, $\delta_M$, $R_0$, and $\theta_N$ were modified to ensure that the timescales of infection clearance and homeostatic response were separated enough for us to use the adaptive programming method, as described in Table S1 of the Supplementary Information.
The simulations in panel (b) are generated with the parameters given in Table ~\ref{tab:Params}. 
\label{fig:pVary}}
\end{figure}

\subsection{The adaptive immune response is subject to a trade-off between pathogen clearance and suppressing inflammation}

In the original adaptive immune model by Stromberg and Carlson \cite{StrombergCarlson2006}, the sole function of effector cells was to clear pathogens.
However, the diverse repertoire of effector T cells--- including helper T cells, cytotoxic T cells, and regulatory T cells--- can additionally exhibit anti-inflammatory functions \cite{shanker2010adaptive,kim2007adaptive,maloy2003cd4+}.
Incorporating these features in the IIB model leads to a trade-off between pathogen clearance and inflammation suppression that can be explored quantitatively.

Specifically, in the IIB model a proportion $p$ of effector cells are allocated to pathogen clearance (a responsibility of cytotoxic T cells with the aid of helper T cells \cite{sompayrac1999immune}), while a proportion $1-p$ of effector cells are allocated to the production of anti-inflammatory cytokines (a responsibility of regulatory T cells \cite{maloy2003cd4+}).
These dual functions are presented schematically in Fig.~\ref{fig:schematic} and explicitly in Table~\ref{tab:model_eqns}.



For smaller values of $p$, effector cells are increasingly used to combat inflammation, which delays the onset of chronic inflammation as demonstrated in 
Fig.~\ref{fig:pVary}a.
On the other hand, at smaller values of $p$ more resources are allocated to combat inflammation, so the diminished innate response slows the rate of pathogen clearance.
This is quantified in Fig.~\ref{fig:pVary}b, which plots an ensemble average (across 1000 infection sequences) of the cumulative pathogen load $L_{\ell}$ for each infection event: for the $\ell$th infection,
\begin{equation}
L_{\ell}\equiv\int_{t_\ell}^{t_{\ell+1}} P_{S_{\ell}}(t)\ \text dt,
\end{equation}
where $S_{\ell}$ is the pathogen shape and $t_\ell$ the starting time of the $\ell$th infection event.
This measure was originally introduced in Stromberg and Carlson \cite{StrombergCarlson2006} where it was called the ``loss function,'' and we similarly use it here as a proxy for the damage the pathogen inflicts during each infection event.
In Fig.~\ref{fig:pVary}b, smaller values of $p$ lead to greater cumulative pathogen load for each infection event and vice versa.

Note that for $p$ as large as $0.9$ there is a further drop in $L_{\ell}$ for $\ell\gtrsim 60$ due to the onset of the chronic inflammation steady state in a larger proportion of individuals.
Once the chronic inflammation state is reached, the non-zero activated phagocyte population $N^*$ rapidly responds to infection events, efficiently clears pathogens, and returns to the chronic inflammation steady state.
In this sense, the onset of chronic inflammation \textcolor{black}{in aging immune systems} acts as a protective mechanism that shields the immune response from future pathogen encounters and minimizes the damage that pathogen inflicts.
\blue{Interpreted biologically, it is advantageous for organisms to minimize cumulative pathogen load while also avoiding the early onset of chronic inflammation.
Evolutionarily, these two opposing selection forces should lead to an intermediate optimal $p$ in which effector cells allocate resources both to pathogen clearance and innate suppression.}


\section{Discussion}
%

\subsection{Immunosenescence as an emergent immune response}
Though immunosenescence affects every aging individual, the mechanisms through which it develops are not yet fully understood.
In this paper we demonstrate through quantitative modeling how physiological markers of immunosenescence can arise from the accumulated effect of pathogen encounters.
In particular, clonal expansion and homeostatic maintenance lead to an increase in memory cells and a decrease in naive cells, which are qualitatively consistent with their clinically-observed abundances \cite{fulop2018immunosenescence}.
Accumulated memory cells protect the immune system against previously-encountered pathogens, but the shrinkage in the naive cell pool renders the immune system vulnerable to novel pathogens \blue{and is the key indicator of immune system fragility.}
While a similar mechanism was demonstrated in the model by Stromberg and Carlson \cite{StrombergCarlson2006}, in their study the overspecialized immune repertoire led to increased cumulative pathogen load. 
In the IIB model, the acquired immune fragility is characterized by a transition to a chronic inflammatory state, and the timing of this transition is highly variable and depends on the infection history.

In addition to this mechanism of imbalanced immunological space, several other immune functions vary with age and could play a role in the development of immunosenescence.
For example, clinical studies observed that the average cytotoxicity of natural killer cells decreases with age \cite{AwPalmer2007, WeinbergerGrubeck2012}, cell signaling between immune cells can become impaired with age \cite{FulopLarbi2014}, and thymus involution leads to decreased T cell production with age \cite{palmer2013effect, lynch2009thymic}.
The current formulation of the IIB model exhibits an inflammaging-like behavior without taking these additional factors into account.
However, in future work physiological parameters of the model could be used as a proxy for these observed behaviors: for example, the decreased cytotoxicity of natural killer cells (which are innate) could be incorporated by decreasing $k_{pn}$ with age, the impaired cell signaling in T cells (which are adaptive) could be achieved by decreasing $\gamma$ with age, or the reduced thymus output could be modeled by decreasing $\theta_N$ with age.
The calculated immune outcomes that result from these modifications could shed light on the relative contributions to immunosenescence from memory-induced fragility in the adaptive response, an impaired innate response, and an impaired adaptive response.

\subsection{\blue{Chronic inflammation as inflammaging and the collaboration between innate and adaptive responses}}
The chronic inflammation steady state has two physiological interpretations.
First, the runaway tissue damage caused by the sustained inflammatory response may cause death in the host, as was implied by Reynolds \textit{et al}.~when they called this steady state ``aseptic death.''
Second, if the sustained inflammatory response is relatively minor, the chronic inflammation state can be interpreted as ``inflammaging,'' a chronic low-grade inflammation that is common among the elderly  \cite{franceschi2000inflamm}.
In this paper we choose this second interpretation and construe the transition to the chronic inflammation state as inflammaging.
\blue{Accordingly, the mechanisms of the IIB model that induce this transition might inform the biological mechanisms that they emulate.}

For example, recent work has suggested that the development of inflammaging might be a result of immune system remodeling: as immunosenescence lessens the efficacy of the adaptive immune response, the body relies on inflammaging for protection against pathogens via the innate immune response \cite{fulop2018immunosenescence}.
Similarly, the adaptive response in the IIB model is subject to a trade-off between clearing pathogens and suppressing inflammation.
In part based on recent work demonstrating that the adaptive response can act to suppress a hazardous innate response \cite{kim2007adaptive}, recent theories suggest that this suppression might have been the evolutionary driver that promoted the development of an adaptive immune response \cite{palm2007not}.
Correspondingly, in the IIB model when pathogens are introduced to a system in the chronic inflammation state, they are cleared almost immediately since the inflammatory response is already primed.

Evolutionarily, the innate immune response preceded the creation of the adaptive response \cite{Bayne2003}.
This is consistent with the taxonomic complexity of organisms, in which invertebrates possess only an innate response while vertebrates possess the additional capacity for pathogen-specific immune memory \cite{MullerPawelec2013}.
Additionally, adaptive immune components are dependent on innate cells--- for example, the activation of an adaptive response through antigen presentation relies on dendritic cells.
\blue{The evolutionary drivers of the adaptive immune response could be explored with immune models that quantify the added benefit of possessing an adaptive immune system.}

\subsection{Age-dependent strength of immune response}
The efficiency of the human immune system changes in a non-monotonic manner as one ages: it is weak in infancy and dependent on maternal antibodies; then it grows stronger as the innate and adaptive responses mature and as immune memory is accumulated; and finally it plummets in the elderly \cite{simon2015evolution}.
\blue{As people age, effector T cell levels drop, chronic inflammation builds \cite{wikby2006immune, whiting2015large}, and immune outcomes among the elderly become extremely variable \cite{whiting2015large}.}

In this work we present a potential mechanism for these clinically-observed aging trends, driven by overspecialization of the adaptive immune repertoire. 
The accumulation of memory cells initially strengthens the immune response against previously-encountered pathogens. 
Eventually, memory cells become overspecialized and restrict the growth of naive cells, rendering aged individuals vulnerable to rare pathogen types.
In the IIB model the onset of the chronic inflammation state is variable, and dependent on the history of previous pathogen encounters.
The age-dependent immune system efficiency observed in the IIB model is consistent with the previously mentioned clinically-observed immune behaviors.

\subsection{Imprinting and vaccines}
The shape space formulation of the adaptive immune response produces results that are qualitatively similar to the clinically-observed behaviors of immune imprinting \cite{GosticLloydSmith2016} and the decreased efficacy of vaccines in the elderly \cite{WeinbergerGrubeck2012}.
Immune imprinting occurs when individuals exhibit sustained memory to the pathogens they were exposed to early in their life.
In the IIB model naive cells are more abundant at the beginning of an infection sequence, and as memory cells accumulate over time, homeostatic pressures drive down the population of naive cells.
Thus, during the first several infection events the larger naive cell pool will induce a stronger adaptive response and therefore generate a stronger memory for encountered pathogens.
On the contrary, near the end of an infection sequence the diminished naive pool will induce a weaker adaptive response to a novel pathogen, and generate a weaker immune memory.
If we interpret vaccination as an exposure to a novel pathogen, then the clinically-observed characteristics of immune imprinting and vaccination in the elderly are qualitatively captured by the IIB model.

\section{Conclusion}
The progression towards immunosenescence is a dynamical process influenced by a lifetime of pathogen encounters, physiological alterations, genetic factors, and general lifestyle choices.
This blend of factors makes it difficult to isolate and identify the most relevant causative agents of immunosenescence. 
Therefore, mathematical models hold great utility in their ability to probe the mechanisms of immunosenescence.

In this paper we developed the IIB model, which incorporates the structure of the innate and adaptive immune branches, and exhibits behaviors that are qualitatively consistent with clinically-observed phenomena.
\blue{We found that repeated pathogen encounters cause an overspecialization of memory cells and depletion of naive cells \textcolor{black}{as the immune system ages}. 
Over time these effects render the immune system \textcolor{black}{fragile} to novel pathogens, the encounter of which will \textcolor{black}{trigger an irreversible transition of the system to a chronic inflammation state}. 
By describing immune dynamics with a mathematical model, we demonstrated how the feedback between innate and adaptive immune responses could give rise to diverse immune courses and outcomes.
Going forward, experimental studies combined with knowledge-based
quantitative models will continue to illuminate the impact of aging on immune efficacy.
}

\section{Acknowledgements}
We thank Andrea Graham, Micaela Martinez, and members of the Aging and Adaptation in Infectious Diseases working group at the Santa Fe Institute for thoughtful discussions regarding the formulation of the IIB model.
This publication is based on work supported by the Santa Fe Institute through the Complex Time: Adaptation, Aging, Arrow of Time research theme, which is funded by the James S. McDonnell Foundation (Grant No. 220020491);
by the National Science Foundation Graduate Research Fellowship under Grant No. 1650114; and by the David and Lucile Packard Foundation and the Institute for Collaborative Biotechnologies through grant W911NF-09-0001 from the U.S. Army Research Office; and by funding from the Chancellor's Office of UCLA.
\newpage

\end{document}